\def\rn{\noindent\parshape 2 0truecm 8.5truecm 0.3truecm 8.2truecm}
\def\rn{}
\def\nn#1 #2{#2. #1}				
\def\nnn#1 #2 #3{#2. #3. #1}			
\def\nnnn#1 #2 #3 #4{#2. #3. #4. #1}		
\def\nnnnn#1 #2 #3 #4 #5{#2. #3. #4. #5. #1}	
\def\dualand{ and\hbox{ }}				
\def\multiand{, and\hbox{ }}				
\def\rf#1;#2;#3;#4;#5 {{\frenchspacing\par\rn#1, #3 {\bf #4}, #5 (#2). \par}}
\def\rg#1;#2;#3;#4;#5;#6 {{\frenchspacing\par\rn#1, #3 {\bf #4}, #5 (#2). \par}}
\def\rfbook#1;#2;#3;#4;#5 {{\frenchspacing\par\rn#1, {\it #3} (#5, #4, #2).\par}}
\def\rfprep#1;#2;#3 {{\par\frenchspacing\rn#1, #3 (#2).\par}}

\def\Mpc{{\rm Mpc}}

\def\expec#1{\langle#1\rangle}

\def\etal{{\frenchspacing\it et al.}}
\def\ie{{\frenchspacing\it i.e.}}
\def\eg{{\frenchspacing\it e.g.}}
\def\etc{{\frenchspacing\it etc.}}

\def\beq#1{\begin{equation}\label{#1}}
\def\eeq{\end{equation}}
\def\beqa#1{\begin{eqnarray}\label{#1}}
\def\eeqa{\end{eqnarray}}
\def\eq#1{equation~(\ref{#1})}
\def\Eq#1{Equation~(\ref{#1})}

\def\fig#1{Figure~\ref{#1}}
\def\Fig#1{Figure~\ref{#1}}

\def\sec#1{Section~\ref{#1}}

\def\spose#1{\hbox to 0pt{#1\hss}}
\def\simlt{\mathrel{\spose{\lower 3pt\hbox{$\mathchar"218$}}
     \raise 2.0pt\hbox{$\mathchar"13C$}}}
\def\simgt{\mathrel{\spose{\lower 3pt\hbox{$\mathchar"218$}}
     \raise 2.0pt\hbox{$\mathchar"13E$}}}
\def\simpropto{\mathrel{\spose{\lower 3pt\hbox{$\mathchar"218$}}
     \raise 2.0pt\hbox{$\propto$}}}

\def\ed{\end{document}}

\def\Pstar{P_*}
\def\Pstarhat{\widehat{\Pstar}}
\def\Phat{\widehat{P}}
\def\Pfid{P_{\rm fid}}
\def\dfid{d_{\rm fid}}
\def\Mapp{M_{\rm app}}
\def\Pnl{P_{\rm nl}}
\def\PPnl{\P^{\rm nl}}
\def\Pl{\P^{\rm l}}
\def\Pz{\P^{\rm z}}
\def\knl{k_{\rm nl}}
\def\keff{k_{\rm eff}}

\def\Deltanl{\Delta_{\rm nl}}

\def\C{{\cal C}}
\def\P{{\cal P}}

\def\l{\ell}

\def\m{{\bf m}}
\def\n{{\bf n}}
\def\p{{\bf p}}

\def\x{{\bf x}}
\def\y{{\bf y}}
\def\z{{\bf z}}

\def\bzero{{\bf 0}}

\def\F{{\bf F}}

\def\I{{\bf I}}
\def\L{{\bf L}}

\def\N{{\bf N}}
\def\Q{{\bf Q}}

\def\NN{{\bf\Sigma}}

\def\W{{\bf W}}
\def\tr{\hbox{tr}\,}

\def\ith{i^{th}}

\def\Ob{\Omega_{\rm b}}
\def\Oc{\Omega_{\rm cdm}}
\def\Od{\Omega_{\rm dm}}
\def\Ok{\Omega_{\rm k}}
\def\Ol{\Omega_\Lambda}
\def\Om{\Omega_{\rm m}}
\def\On{\Omega_\nu}
\def\Ox{\Omega_x}
\def\ob{\omega_{\rm b}}
\def\ocdm{\omega_{\rm cdm}}
\def\od{\omega_{\rm dm}}

\def\om{\omega_{\rm m}}

\def\fn{f_\nu}

\def\ns{n_s}
\def\nt{n_t}
\def\As{A_s}
\def\At{A_t}

\def\L{{\cal L}}

\def\l{\ell}

\def\ith{i^{th}}

\documentstyle[prd,aps,epsf]{revtex}
\begin{document}
\twocolumn[\hsize\textwidth\columnwidth\hsize\csname@twocolumnfalse\endcsname



\title{Separating the Early Universe from the Late Universe:\\
cosmological parameter estimation beyond the black box}

\author{Max Tegmark$^1$ \& Matias Zaldarriaga$^{2,3}$}

\address{$^1$Dept. of Physics, Univ. of Pennsylvania, Philadelphia, PA 19104;
  max@physics.upenn.edu}

\address{$^2$Dept. of Physics, New York University,  
4 Washington Pl., New York, NY 10003;mz31@nyu.edu}

\address{$^3$Institute for Advanced Study, Einstein Drive, Princeton NJ 08540}

\date{Submitted to {\it Phys. Rev. D} Jul 3 2002, accepted Aug 26 2002 }

\maketitle 

\begin{abstract}
We present a method for measuring the cosmic matter budget without
assumptions about speculative Early Universe physics,
and for measuring the primordial power spectrum 
$\Pstar(k)$ non-parametrically,
either by combining CMB and LSS information or by using CMB polarization.
Our method complements
currently fashionable ``black box'' cosmological
parameter analysis, constraining cosmological models 
in a more physically intuitive fashion by mapping
measurements of CMB, weak lensing and cluster abundance
into $k$-space, where they can be directly compared with each other and with
galaxy and Ly$\alpha$ forest clustering.
Including the new CBI results, we find that CMB measurements
of $P(k)$ overlap with those from 2dF galaxy clustering 
by over an order of magnitude in scale, and even overlap with 
weak lensing measurements.
We describe how our approach can be used to raise the ambition
level beyond cosmological parameter fitting as data improves,
testing rather than assuming the underlying physics. 

\end{abstract}
\bigskip
] 


\section{Introduction} 

What next? 
An avalanche of measurements have now lent support to a 
cosmological ``concordance model'' whose free parameters have
been approximately measured, tentatively answering 
many of the key questions posed in past papers.
Yet the data avalanche is showing no sign of abating, with spectacular
new measurements of the cosmic microwave background (CMB), 
galaxy clustering, Lyman $\alpha$ forest (Ly$\alpha$F) clustering and
weak lensing expected in coming years.
It is evident that many scientists, despite putting on a brave face, 
wonder why they should care about all this new data if they already know the basic
answer.
The awesome statistical power of this new data can be used in two ways:
\begin{enumerate}
\item To measure the cosmological parameters of the concordance model
(or a replacement model if it fails) to additional decimal places
\item To test rather than assume the underlying physics
\end{enumerate}
This paper is focused on the second approach, which has received less
attention than the first in recent years.
As we all know, cosmology is littered with ``precision'' measurements that came and went.
David Schramm used to hail Bishop Ussher's calculation
that the Universe was created 4003 b.c.e. as a fine example --- small statistical
errors but potentially large systematic errors.
A striking conclusion from comparing recent parameter estimation papers
(say \cite{9par,10par,concordance,consistent} by the authors for
methodologically uniform sample) is that the quoted error bars have not
really become smaller, merely more believable.
For instance, a confidence interval for the dark energy density
that would be quoted three years ago by
assuming that four disparate data sets were all correct \cite{9par} can now be
derived from CMB + LSS power spectra alone \cite{consistent,Efstathiou02,Melchiorri02,Lewis02} and independently
from CMB + SN 1a as a cross-check.

\begin{figure}[tb] 
\centerline{\epsfxsize=9.0cm\epsffile{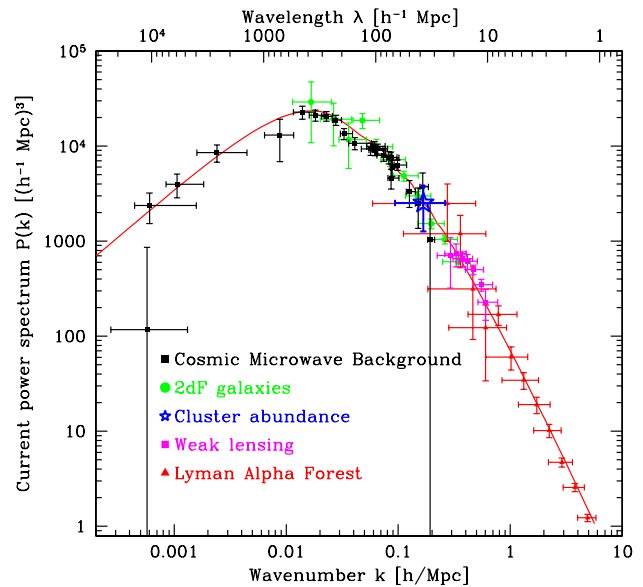}}
\vskip-0.8cm
\smallskip
\caption{\label{pplotFig}\footnotesize%
Measurements of the linear matter power spectrum $P(k)$ computed as described in the text,
using the concordance model of \protect\cite{Efstathiou02} (solid curve) to compute window functions.
The locations of the CMB points depend on the matter budget and scales with the reionization
optical depth as $e^{2\tau}$ for $k\simgt 0.002$. Correcting for bias 
shifts the 2dF galaxy points 
\protect\cite{2df} vertically ($b=1.3$ assumed here)
and should perhaps blue-tilt them slightly. 
The cluster point scales vertically as $(\Omega_m/0.3)^{-1.2}$, and its
error bars reflects the spread in the literature.
The lensing points are based on \protect\cite{Hoekstra02}. 
The Ly$\alpha$F points are from a reanalysis 
\protect\cite{GnedinHamilton01} of \protect\cite{Croft00} and have an overall calibration 
uncertainty around 17\%.
}
\end{figure}

This paper aims to extend this trend, showing how measurements can be combined
to raise the ambition level beyond simple
parameter fitting, testing rather than assuming the underlying physics.
Many of the dozen or so currently fashionable cosmological parameters 
merely parametrize two cosmological functions \cite{gravity,spacetime}:
the cosmic expansion history $a(t)$ and the cosmic clustering history
$P(k,z)$, the observables corresponding to 0th and 1st order cosmic perturbation theory,
respectively.  
This means that non-parametric measurements of these cosmological functions
allows testing whether the assumptions associated with the
cosmological parameters are in fact valid. Moreover, if there are discrepancies, 
comparing measurements of these functions from different data sets reveals whether
the blame lies with theory, data or both. 

We will limit our treatment to the 1st order function, $P(k,z)$, since the 0th order function
has been extensively discussed previously \cite{gravity,spacetime,WangGarnavich01}.
One of the key ideas of this paper is 
summarized in \fig{pplotFig}, showing 
how CMB, LSS, clusters, weak lensing and Ly$\alpha$F all constrain 
$P(k,z)$ at $z=0$. The first plots that we are aware of showing CMB
in k-space go back a decade \cite{ScottWhiteSilk95},
when CMB merely probed scales much larger than accessible to large-scale structure 
measurements. Since then, CMB has gradually pushed to smaller scales with 
improved angular resolution while LSS has pushed to larger scales with deeper 
galaxy surveys. What is particularly exciting now, and makes this paper timely, is
that the two have met and overlapped, especially with the CBI experiment \cite{CBI} 
and the 2dF \cite{Colless01} and SDSS \cite{York00} redshift surveys. \Fig{pplotFig}
shows that CMB now overlaps also with the scales probed by cluster abundance and even,
partly, with weak lensing.
 
$P(k,z)$ can be factored as the product of a primordial power spectrum 
$\Pstar(k)$ and a transfer function, corresponding to the physics of the 
Early Universe and the Late Universe, respectively\footnote{We will assume that the
primordial fluctuations are adiabatic, discussing the most general case
in \sec{DiscSec}.}.
The two involve completely separate physical processes
and assumptions that need to be tested, and the purpose of our method is to  
measure these two factors separately using observational data.
Given a handful of cosmological parameters specifying the cosmic matter
budget and the reionization epoch,
the transfer function can be computed from first principles using 
well-tested physics (linearized gravity and plasma physics at temperatures 
similar to those at the solar surface).
The primordial power spectrum is on shakier ground, generally believed to 
have been created in the Early Universe at an energy scale never observed 
and involving speculative new physical entities.
Most work has parametrized this function as
a power law $\Pstar(k)\propto k^n$ or a logarithmic parabola
$\Pstar(k)\propto k^{n+\alpha \ln k}$, inspired by the slow-roll approximation in inflationary
models \cite{LiddleBook}, usually with $\alpha=0$.
More general parametrizations have included broken power 
laws \cite{Amendola95,Kates95,Atrio97,Einasto99} a piecewise constant 
function \cite{Wang99} and other forms \cite{Kinney01,Matsumiya02}. 
It has also been shown \cite{Wang99} that 
the MAP CMB data \cite{MAP} in combination with SDSS power spectrum measurements
should be able to constrain the shape of $\Pstar(k)$ in considerable detail.
The key challenge is breaking the degeneracy between the two factors,
$\Pstar(k)$ and the transfer function.
Although a future brute-force likelihood analysis parametrizing
$\Pstar(k)$ with, say, 20 parameters would be interesting and perfectly valid,
 it would obscure the simplicity of the underlying physics. 
Such a ``black box'' approach would entail computing many different curves
for each point in parameter space (such as $C_\l$ for CMB, $P(k)$ for galaxies, 
the aperture mass function $\Mapp(\theta)$ for lensing and 
the cluster mass function), and mapping out the 20-dimensional likelihood function 
numerically by marginalizing over other cosmological parameters like those of the
matter budget. This would be overkill, since (modulo nonlinearity complications
treated below) all measurements shown in \fig{pplotFig} can be recast directly as 
weighted averages of $\Pstar(k)$.

The rest of this paper is organized as follows.
In \sec{Psec}, we describe the construction of \fig{pplotFig}, explaining how 
CMB, weak lensing and cluster abundance measurements can be mapped 
into (linear) $k$-space.
In \sec{DegenSec}, we turn to the degeneracy between $\Pstar(k)$ and
cosmological parameters such as the various matter densities, and
present our method for breaking it. We show how this allows measuring the 
cosmic matter budget without assuming anything about $\Pstar(k)$
and obtaining a non-parametric measurement of $\Pstar(k)$.

\def\pk{P(k)}  
\section{Measuring $\pk$ when the transfer functions are known}
\label{Psec}

In this section, we discuss how measurements of CMB, weak lensing, cluster abundance,
Ly$\alpha$F and galaxy clustering probe $P(k)$ and $\Pstar(k)$
when the relevant transfer functions are known. We will see that in all five cases, each measured data point $d_i$
(a CMB band power $\delta T_\l^2$, a lensing aperture mass 
variance $\expec{\Mapp(\theta)^2}$, \etc)
 can be written as an integral 
\beq{WindowDefEq}
d_i = \int_{-\infty}^\infty \P_i(k)d\ln k
\eeq
over (linear) wavenumber $k$ for some non-negative integrand $\P_i(k)$.
Renormalizing $\P_i(k)$ to be a probability distribution, 
our convention $(\star)$ in \fig{pplotFig} is the following:
\begin{itemize}
\item[($\star$)] {\it Plot the data point at the
$k$-value corresponding to the median of this distribution with a horizontal bar
ranging from the 20th to the 80th percentile.}
\end{itemize}
These percentiles correspond to the full-width-half-maximum for the special case of a
Gaussian distribution. In other words,
the horizontal bars indicate the range of scales $k$ contributing to the data point.
All transfer functions in this section are computed assuming the flat $\Lambda$CDM
concordance model of \cite{Efstathiou02} --- we return the more general case in the next section.
This is a flat, scalar, scale-invariant model with
cold dark matter density  $h^2\Omega_m=0.12$, baryon density $h^2\Omega_b=0.021$
and cosmological constant $\Ol=0.71$. This corresponds to a 
Hubble parameter $h = 0.70$. 
We chose a reionization optical depth $\tau=0.05$, which corresponds to reionization
at redshift $z=8$ with these parameters. 
We normalize the model to have $\sigma_8=0.815$, which provides a good fit to the CMB data.

\subsection{CMB data}

\begin{figure}[tb] 
\vskip-1.2cm
\centerline{\epsfxsize=9.0cm\epsffile{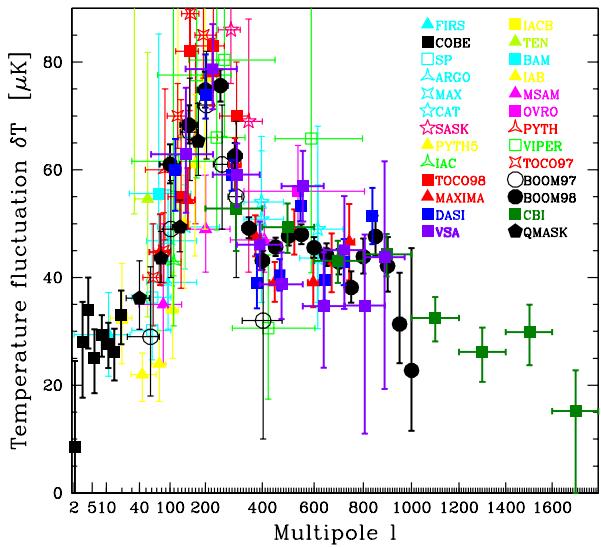}}
\vskip-0.8cm
\smallskip
\caption{\label{cmb_experimentsFig}\footnotesize%
CMB data used in our analysis.
Error bars in the plot do not include calibration or beam errors
which allow substantial vertical shifting and 
tilting for some experiments (these effects were included in our
analysis). 
}
\end{figure}

\Fig{cmb_experimentsFig} shows all 119 CMB detections currently available, 
extending the compilation in \cite{consistent} by adding 
the new measurements from the Very Small Array (VSA) \cite{VSA} and the
Cosmic Background Imager (CBI) mosaic \cite{CBI}. 
Recent data reviews include \cite{Gawiser00,radpack}.

\begin{figure}[tb] 
\vskip-1.2cm
\centerline{\epsfxsize=9.0cm\epsffile{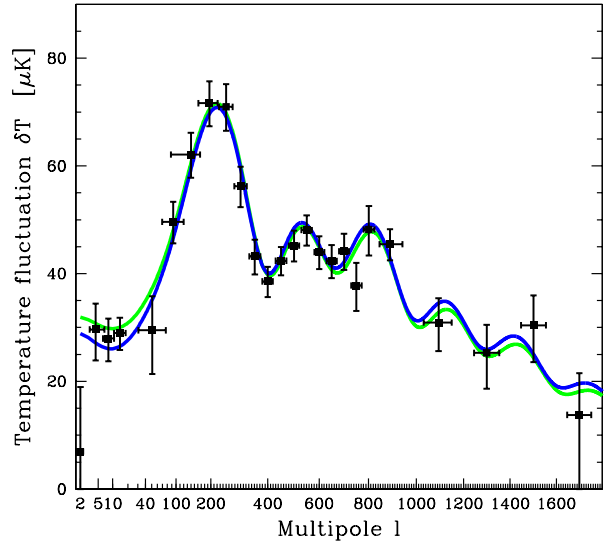}}
\vskip-0.8cm
\smallskip
\caption{\label{cmb_combinedFig}\footnotesize%
Combination of data from \protect\fig{cmb_experimentsFig}.
These error bars include the effects of beam and
calibration uncertainties, which cause long-range correlations
of order 5\%-10\% over the peaks. In addition, points tend to be anti-correlated 
with their nearest neighbors, typically at the level of 
a few percent.
The horizontal bars give the characteristic widths of the window functions (see text).
The curves show the flat $\Lambda$CDM  concordance models from
\protect\cite{consistent} (green/light gray) and \protect\cite{Efstathiou02} (blue/dark gray).
}
\end{figure}

We combine these measurements into a single set of 25 band powers shown in
\fig{cmb_combinedFig} and Table 1 using
the method of \cite{consistent}, including calibration and beam 
uncertainties, which effectively calibrates the experiments against each other. 
The coefficients $y_i$ in equations (A2), (A4) and (A5) of \cite{consistent}
should strictly speaking be the window-convolved true power spectrum, so rather than
approximating them by the observed data points as in \cite{consistent}, we 
approximate them by the 
the smooth fit to the data given by the green/light gray curve in \fig{cmb_combinedFig} 
convolved with the experimental window functions.
We have excluded the PythV data since it disagrees with numerous 
other experiments on large angular scales and the non-release of the underlying 
map precludes clarifying this situation.
Since our compressed band powers $d_\l$ are simply linear combinations 
of the original measurements, they can be
analyzed ignoring the details of how they were constructed, being 
completely characterized by a window 
matrix $\W$:
\beq{lwindowEq}
\expec{d_i} = \sum_{\l}\W_{i\l}\C_\l,
\eeq
where $\C_\l\equiv\delta T_\l^2\equiv\l(\l+1)C_\l/2\pi$ and 
$C_\l$ is the angular power spectrum.
This matrix is available at $www.hep.upenn.edu/{\sim}max/cmb/experiments.html$
together with the 25 band powers $d_\l$ and their 
$25\times 25$ covariance matrix.
Following the convention used in \fig{pplotFig}, 
the data $\l$-values and effective $\l$-ranges in \fig{cmb_combinedFig} and Table 1
correspond to the median, 20th and 80th percentile of the window functions $\W$.
(We use absolute values of the window function to be pedantic, since 
some windows go slightly negative 
in places as a result of the inversion, although this makes a negligible difference for the plot.)
Comparing Table 1 with the older results from \cite{consistent}, we find that the degree-scale
normalization is marginally higher. In bin 8, for instance, corresponding to the 1st peak, the
normalization has risen by 3\% due to the inclusion of the VSA and CBI results
and a further 6\% due to the above-mentioned improved modeling of calibration and beam 
errors. With the old modeling, a measurement scattering low by chance would be assigned 
a smaller error and therefore get more statistical weight, pulling the overall calibration
down somewhat. A detailed discussion of calibration issues can be found in \cite{Bridle01}.

\bigskip
\bigskip
\bigskip

\bigskip
\noindent
{\footnotesize
{\bf Table 1} -- Band powers combining the
information from CMB data 
from \protect\fig{cmb_experimentsFig}.
The 1st column gives the $\l$-bins used when combining the data, and can be ignored
when interpreting the results.
The 2nd column gives the medians and characteristic widths of 
the window functions as detailed in the text. 
The error bars in the 3rd column
include the effects of calibration and beam uncertainty.
The full $25\times 25$ correlation matrix and $25\times 2000$
window matrix are available at $www.hep.upenn.edu/\sim max/cmb/experiments.html$.
\bigskip
\begin{center}
{\footnotesize
\begin{tabular}{|c|c|c|}
\hline
$\l$-Band	&$\l$-window	&$\delta T^2\>[\mu $K$^2]$\\
\hline		
$    2-   2$	 &$    2_{- 0}^{+ 0}$&$  49\pm 310$\\
$    3-   5$	 &$    4_{- 1}^{+ 2}$&$ 878\pm 308$\\
$    6-  10$	 &$    8_{- 2}^{+ 3}$&$ 782\pm 219$\\
$   11-  30$	 &$   14_{- 3}^{+ 4}$&$ 840\pm 171$\\
$   31-  75$	 &$   51_{-20}^{+26}$&$ 869\pm 412$\\
$   76- 125$	 &$   93_{-24}^{+26}$&$2457\pm 383$\\
$  126- 175$	 &$  140_{-52}^{+26}$&$3854\pm 520$\\
$  176- 225$	 &$  195_{-34}^{+26}$&$5130\pm 595$\\
$  226- 275$	 &$  250_{-24}^{+23}$&$5036\pm 613$\\
$  276- 325$	 &$  300_{-22}^{+23}$&$3163\pm 423$\\
$  326- 375$	 &$  352_{-20}^{+21}$&$1869\pm 278$\\
$  376- 425$	 &$  399_{-20}^{+20}$&$1486\pm 216$\\
$  426- 475$	 &$  451_{-20}^{+21}$&$1795\pm 221$\\
$  476- 525$	 &$  500_{-20}^{+21}$&$2043\pm 260$\\
$  526- 575$	 &$  550_{-21}^{+21}$&$2312\pm 271$\\
$  576- 625$	 &$  601_{-20}^{+21}$&$1936\pm 269$\\
$  626- 675$	 &$  650_{-21}^{+21}$&$1792\pm 261$\\
$  676- 725$	 &$  700_{-20}^{+20}$&$1950\pm 295$\\
$  726- 775$	 &$  750_{-21}^{+22}$&$1429\pm 336$\\
$  776- 825$	 &$  802_{-21}^{+23}$&$2322\pm 440$\\
$  826-1000$	 &$  890_{-44}^{+51}$&$2065\pm 261$\\
$ 1001-1200$	 &$ 1094_{-63}^{+56}$&$ 955\pm 300$\\
$ 1201-1400$	 &$ 1299_{-54}^{+54}$&$ 639\pm 291$\\
$ 1401-1600$     &$ 1501_{-54}^{+53}$&$ 925\pm 368$\\
$ 1601-\infty$   &$ 1700_{-53}^{+51}$&$ 190\pm 273$\\
\hline		
\end{tabular}
}
\end{center}
}


The angular power spectrum of the CMB is determined by the primordial power
spectrum $\Pstar(k)$ through a linear relation
\beq{CMBwindowEq}
\C_\l = \int_{-\infty}^\infty W_\l(k)\Pstar(k)d\ln k,
\eeq
where the transfer functions $W_\l(k)$ depend on the cosmic matter budget
and the reionization optical depth.
Since
\beq{MatterTransferEq}
P(k)=T(k)^2\Pstar(k)
\eeq
where $T(k)$ is the matter transfer function, \eq{CMBwindowEq} can be reexpressed directly in
terms of the current power spectrum:
\beq{CMBwindowEq2}
\C_\l = \int_{-\infty}^\infty {W_\l(k)\over T(k)^2} P(k)d\ln k.
\eeq

\begin{figure}[tb] 
\vskip-1.2cm
\centerline{\epsfxsize=9.0cm\epsffile{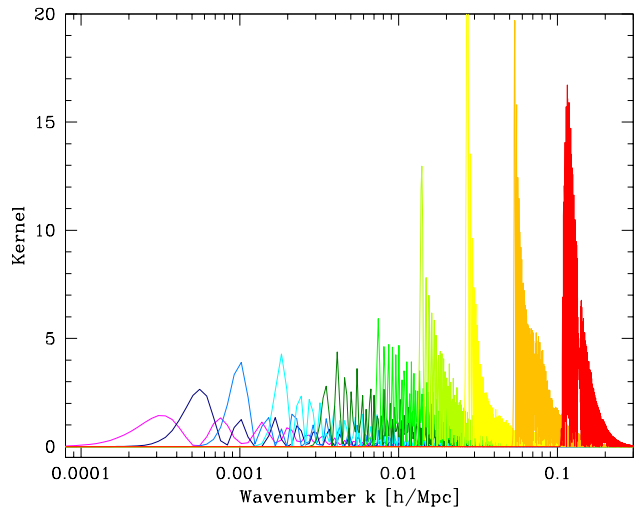}}
\vskip-0.8cm
\smallskip
\caption{\label{kernelsFig}\footnotesize%
The curves $k W_\l(k)$ whose integral give 
$\C_\l$ for a scale-invariant spectrum, all rescaled to have unit area. 
From left to right, the curves are for multipoles
$\l=2$, 4, 8, 16, 32, 64, 128, 256, 512, 1024.
}
\end{figure}

\begin{figure}[tb] 
\centerline{\epsfxsize=12.5cm\epsffile{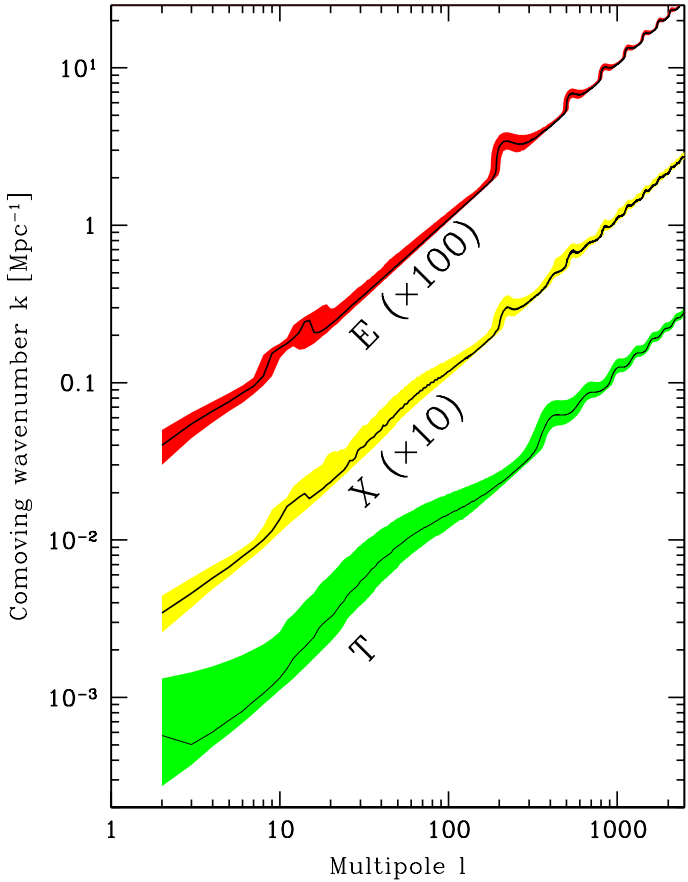}}
\smallskip
\caption{\label{keffFig}\footnotesize%
The correspondence between $\l$-space and $k$-space for CMB.
For each $\l$, the shaded bands indicates the $k$-range from the 20th to 80th percentile 
of the distribution $k W_\l(k)$ (\fig{kernelsFig}), and the black curve shows the median.
From top to bottom, the three bands are for the E-polarization, cross-polarization (X)
and unpolarized (T) cases, respectively. To avoid clutter, the E and X bands have been multiplied by
10 and 100, respectively.
}
\end{figure}

\Eq{CMBwindowEq} implies that for the CMB with data points $d_\l\equiv\C_\l$, 
the integrand of \eq{WindowDefEq} is simply
$\P_\l(k) = W_\l(k)\Pstar(k)$.
We compute the integral kernels $W_\l(k)$ with CMBfast \cite{cmbfast}. \Fig{kernelsFig} shows
$\P_\l(k)$ for a sample of $\l$-values, normalized to integrate to unity, for a scale-invariant 
primordial power spectrum $\Pstar(k)\propto k$. In other words, the
figure simply shows $k W_\l(k)$ rescaled, the integral of which gives 
$\C_\l$. For each such curve, we compute the 20th, 50th and 80th percentile as per the
above-mentioned convention $(\star)$ and 
plot the results in \fig{keffFig} as an indication of what $k$-range is probed by each 
multipole $\l$. The situation is completely analogous for the polarization case.
The relations between $\l$ and $k$ are seen to be roughly linear
as expected, and to tighten with increasing $\l$. 


The slight wiggles roughly line up with 
the derivatives of the three CMB power spectra. This is because when the power spectrum is 
steeply rising, the contribution will be larger from the peak on the right than from the trough on 
the left, pushing the median up towards higher $k$, and vice versa.
These wiggles are seen to be more pronounced for E-polarization than for the unpolarized
case. This is because the wiggles are sharper and have greater relative amplitude for 
the polarized case, increasing the magnitude of the derivative. The T-spectrum has milder wiggles
since the peculiar velocity contribution fills in the troughs between the peaks from
the dominant density/gravitational contribution --- the E-power spectrum has only a velocity 
contribution and thus drops near zero between peaks, staying positive only because of
geometric projection effects in the mapping from $k$-space to $\l$-space.

\begin{figure}[tb] 
\vskip-1.2cm
\centerline{\epsfxsize=9.0cm\epsffile{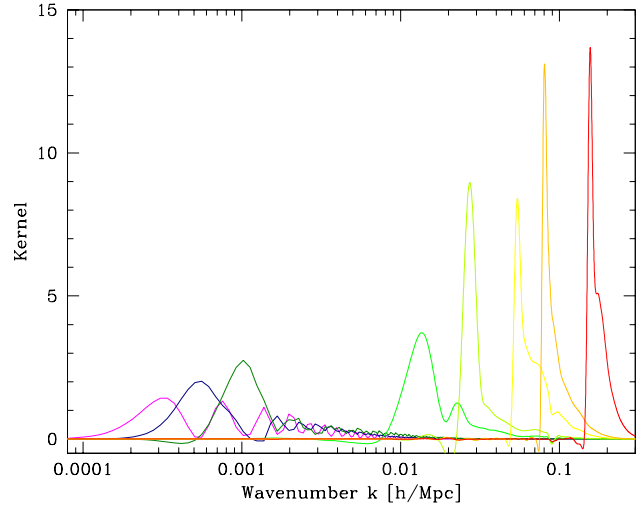}}
\vskip-0.8cm
\smallskip
\caption{\label{cmb_windowsFig}\footnotesize%
Sample curves $k W_i(k)$ whose integral give 
our binned CMB  band powers from \fig{cmb_combinedFig} for a scale-invariant spectrum, 
all rescaled to have unit area. 
From left to right, the curves correspond to band powers  
$i=1$,  2, 3, 6, 9, 14, 19, 24.
}
\end{figure}

Because of incomplete sky coverage, real-world CMB measurement can never measure 
individual multipoles $\l$, merely weighted averages of many. Substituting 
\eq{CMBwindowEq} into \eq{lwindowEq} gives 
\beq{CMBwindowEq3}
\expec{d_i} = \int_{-\infty}^\infty W_i(k)\Pstar(k)d\ln k,
\eeq
where
\beq{CMBwindowEq4}
W_i(k)\equiv \sum_{\l}\W_{i\l} W_\l(k).
\eeq
In other words, each of our 25 binned CMB measurements probes a known linear
combination of the primordial power spectrum. A sample of these new window functions
are plotted in \fig{cmb_windowsFig}, and are again seen to be quite narrow for large $\l$. 
Indeed, although the $\l$-smearing makes these windows slightly broader than those
in \fig{kernelsFig}, it is also seen to make them more well-behaved, eliminating the
high-frequency oscillations at large $\l$.

\begin{figure}[tb] 
\centerline{\epsfxsize=9.0cm\epsffile{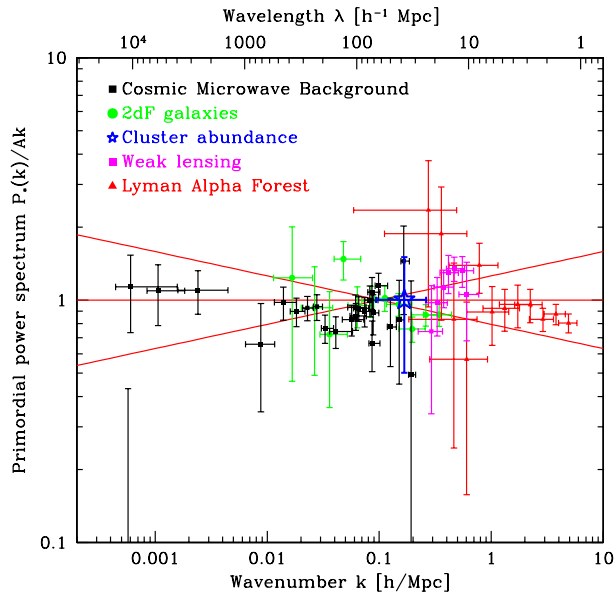}}
\vskip-0.8cm
\smallskip
\caption{\label{kplotFig}\footnotesize%
Measurements of the primordial power spectrum $\Pstar(k)$ computed as described in the text,
divided by our fiducial primordial power spectrum $Ak$ with
$A\approx (43.2\Mpc/h)^4$. This normalization corresponds to an rms
fluctuation in the gravitational potential $\psi$ given by 
$[k^3 P_\psi(k)/2\pi^2]^{1/2}=2.9\times 10^{-5}$.
The data are the same as in \fig{pplotFig}, divided by $A k T(k)^2$.
}
\end{figure}

\begin{figure}[tb] 
\vskip-1.2cm
\centerline{\epsfxsize=9.0cm\epsffile{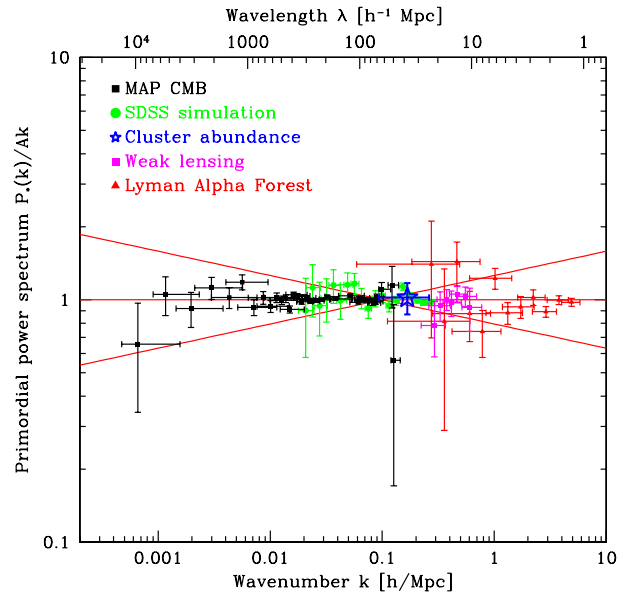}}
\vskip-0.8cm
\smallskip
\caption{\label{kplot_futureFig}\footnotesize%
Same as \fig{kplotFig}, but for simulated future data.
CMB and galaxy points assume the advertized specifications for the complete
MAP and SDSS datasets, respectively. The CMB simulation assumes the MID foreground
model of \protect\cite{foregpars}. Still better galaxy measurements are likely to result 
adding the SDSS luminous red galaxy survey \protect\cite{Eisenstein01} and photometric
redshift information.
The cluster point assumes a relative error of 7.5\% on $\sigma(R)$,
\protect\ie, that systematic uncertainties can be reduced to the level of current statistical errors.
The lensing and Ly$\alpha$F points assume that the current errors have been cut in half.
}
\end{figure}

\begin{figure}[tb] 
\centerline{\epsfxsize=9.0cm\epsffile{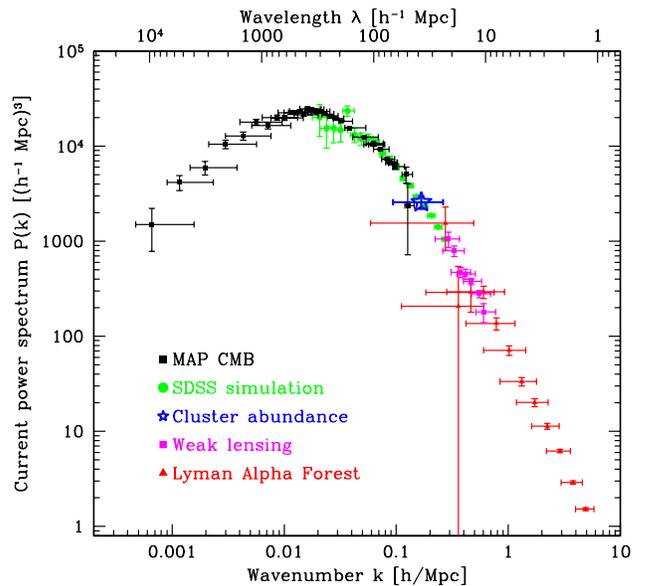}}
\vskip-0.8cm
\smallskip
\caption{\label{pplot_futureFig}\footnotesize%
Same as \fig{kplot_futureFig}, but for the current linear power spectrum $P(k)$.
In other words, this is \fig{pplotFig} for simulated future data.
}
\end{figure}

We are now ready to map our CMB measurements from \fig{cmb_combinedFig} into 
$k$-space. We need a prescription for where to position the points both 
horizontally and vertically. 
Horizontally, we simply follow the above-mentioned convention $(\star)$
and plot it at the median of 
the distribution $\P_i(k)=kW_i(k)$ from \fig{cmb_windowsFig}, with horizontal
bars extending from the 20th to the 80 percentile. 
Vertically, we plot it at the value $\Pstarhat_i$ defined by
\beq{PstarhatEq}
k^{-1}\Pstarhat_i\equiv  {d_i\over \int_{-\infty}^\infty k W_i(k)d\ln k}.
\eeq
Taking the expectation value of this and using \eq{CMBwindowEq3} tells us that we can 
interpret $\Pstarhat_i$ as measuring simply a weighted average of 
$k^{-1}\Pstar(k)$ (which we expect to be a nearly constant function), 
with the window function $k W_i(k)$ giving the weights.
The resulting 25 measurements of $\Pstar(k)$ are shown in \fig{kplotFig}, and 
\fig{kplot_futureFig} shows a simulation for measurements by the MAP 
satellite\cite{MAP}. 
To plot these points as measurements of $P(k)$, we proceed analogously. 
We use exactly the same convention ($\star$) for the horizontal placement of the points, and given
\eq{MatterTransferEq}
plot them at a vertical position given by 
\beq{PhatEq}
\Phat\equiv  T(\keff)^2\Pstarhat,
\eeq
where $\keff$ is the horizontal location of the point (the median of the window function).
This allows us to interpret $\Phat_i/\Pfid$ as measuring simply a weighted  average of the
relative power $P(k)/\Pfid(k)$, where  $\Pfid(k)\equiv kT(k)^2$ is our fiducial power
spectrum. 
This procedure produces the CMB points plotted in \fig{pplotFig} and \fig{pplot_futureFig}.

\begin{figure}[tb] 
\centerline{\epsfxsize=9.0cm\epsffile{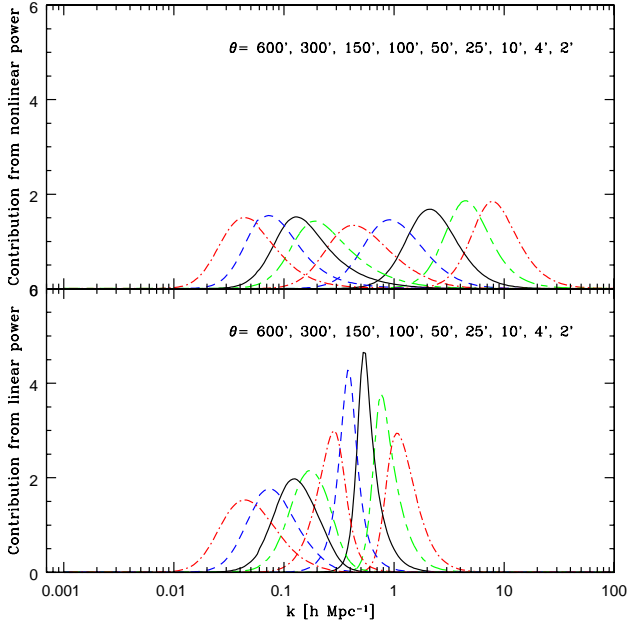}}
\smallskip
\caption{\label{lensing_windows_hoekstraFig}\footnotesize%
Curves show $\PPnl(k)$ (top) and $\Pl$ (bottom), normalized to have unit area.
These curves show how much different nonlinear (top) and linear (bottom) scales contribute 
to the observed weak lensing aperture mass variance $\expec{\Mapp(\theta)^2}$ for the source
redshift distribution of
\protect\cite{Hoekstra02} .
}
\end{figure}

\begin{figure}[tb] 
\centerline{\epsfxsize=9.0cm\epsffile{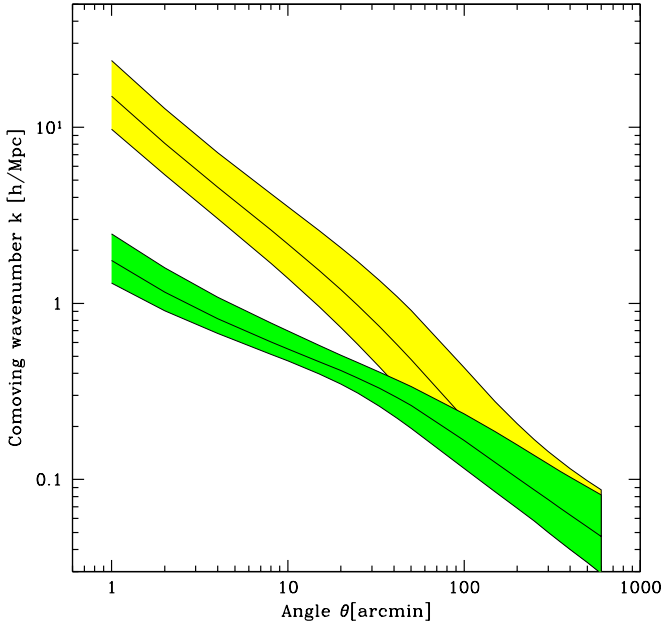}}
\smallskip
\caption{\label{lensing_k_range_hoekstraFig}\footnotesize%
Ranges of $k$ that contribute to $M_{app}(\theta)^2$ as a function of
$\theta$. The line indicates the median $k$ and the boundaries
correspond to the 20th and 80th percentile. 
The upper (lower) band shows the range for nonlinear (linear)
scales.}
\end{figure}

\subsection{Weak lensing data}

Weak gravitational lensing
uses photons from distant galaxies as test particles to
measure the metric fluctuations caused by intervening matter,
as manifested by distorted images.
The first detections of this cosmic shear signal \cite{WeakLensingReview,HoekstraReview}
were reported in 2000
\cite{Wittman00,Waerbeke00,Bacon00,Kaiser00,Rhodes01,Waerbeke01},
and dramatic improvements are likely to lie ahead just as for CMB observations.
For this paper, we will use the results from the Red-Sequence Cluster Survey (RSCS)
reported in \cite{Hoekstra02}, which includes data from a record-breaking 53 square degree
sky area. We use the seven data points employed for the cosmological analysis
in \cite{Hoekstra02}, \ie, 
\beq{LensingDataDef}
d_i\equiv\expec{\Mapp(\theta_i)^2}
\eeq
where $\theta=$10', 15', 20', 25', 35, 33, 42', 50'.
This measured quantity, denoted the aperture mass variance,
is on average given by
\beq{AppMassEq}
\expec{d_i} = 2\pi \int_0^\infty \left[{12 J_4(\l\theta_i)\over\pi(\l\theta_i)^2}\right]^2 \l P_\kappa(\l)d\l,
\eeq
where $J_4$ is a Bessel function and $P_\kappa(\l)$ is the cosmic shear power spectrum.
The shear power spectrum in turn is given by a linear combination of
the nonlinear matter power spectrum $\Pnl(\knl)$ over a range of wavenumbers $\knl$ and 
redshifts \cite{WeakLensingReview,HoekstraReview},
\beq{Pkappa}
P_\kappa(\l)= {9\Omega_m^2 H_0^4\over 4 c^4}\int_0^{\omega_H} 
 \left({\bar W(\omega)\over a(\omega)}\right)^2 \Pnl\left({l\over f_K(\omega)};\omega\right)d\omega.
\eeq
where we have introduced the comoving radial coordinate $\omega$ and
$\omega_H$ 
corresponds to the horizon. Here $f_K(\omega)$ is the
angular diameter distance  and  
$\bar W(\omega)$ is a source-averaged ratio of angular diameter
distances. 
For a given redshift distribution of the sources $p_b(\omega)$,
\beq{barW}
\bar W(\omega)= \int_\omega^{\omega_H} 
p_b(\omega^\prime) 
{f_K(\omega^\prime-\omega) \over f_K(\omega^\prime)}d\omega^\prime.
\eeq
We use the best fit redshift distribution for the RSCS sample from \cite{Hoekstra02},
\beq{HoekstraDistributionEq}
p_b(z) = {\beta(z/z_s)^\alpha\over z_s \Gamma[(1+\alpha)/\beta} 
e^{-(z/z_s)^\beta}
\eeq
with
$(\alpha,\beta,z_s)=(4.7,1.7,0.302)$.
With a  change of variables to $\knl=\l/f_K(\omega)$ 
we can rewrite \eq{AppMassEq} in the form of 
\beqa{LensingWindowDefEq}
d_i &=& \int_{-\infty}^\infty \PPnl_i(\knl)d\ln\knl \nonumber \\
\PPnl_i(\knl)&=& {9\Omega_m^2 H_0^4\over 2 \pi c^4}\int_0^{\omega_H} 
 \left({\bar W(\omega) f_K(\omega)\over a(\omega)}\right)^2 
\knl^2\Pnl\left(\knl;\omega\right)\nonumber \\
&&  \times \left[{12
J_4(\knl f_K(\omega) \theta_i)\over(\knl f_K(\omega) \theta_i)^2}\right]^2
d\omega.
\eeqa
The integrand
$ \PPnl_i(\knl)$ is an integral over cosmic time that depends linearly on the 
nonlinear matter power spectrum $\Pnl(\knl)$ at various redshifts. 
The upper panel in \fig{lensing_windows_hoekstraFig} shows this integrand for 
a sample of angular scales $\theta$.
Just as for the CMB, we follow our convention $(\star)$ and compute
the 20th, 50th and 80th percentiles of these distributions. The results, plotted in
\fig{lensing_k_range_hoekstraFig}, show that we approximately have
$k\propto 1/\theta$ as expected but that the relation is not particularly tight, 
with a given  $\theta$ probing a broad range of $k$-values.

To use  this relation between $k$ and $\theta$ 
to map the lensing data into (linear) $k$-space would be quite misleading,
since the nonlinear power $\Pnl(\knl)$ is the result of gravitational collapse and therefore 
carries information about the linear power on larger spatial scales \cite{HKLM,Jain95,PeacockDodds96}.
We will use the {\it Ansatz} of Peacock \& Dodds \cite{PeacockDodds96} to quantify this effect. 
Defining 
\beq{DeltaDefEq}
\Delta(k) = {4\pi k^3\over (2\pi)^3}P(k),\quad    
\Deltanl(\knl)  = {4\pi \knl^3\over (2\pi)^3}\Pnl(\knl),
\eeq
the linear power $\Delta$ on scale $k$ is approximately related to the nonlinear power
$\Deltanl$ on a smaller nonlinear scale $\knl$, 
\beqa{PeacockDoddsEq}
\Deltanl(\knl)&=&f_{NL}(\Delta(k)),\label{deltanl}\\
\knl&=&(1+\Deltanl)^{1/3} k,\label{knlEq}
\eeqa
where $f_{NL}(x)$ is a fitting function that depends on both the cosmology
and the slope of the linear power spectrum.\footnote{We used a simple
$\Gamma$-model fit to the power spectrum (ie. a fit with no wiggles) 
to calculate the slope needed in the
Peacock and Dodds {\it Ansatz}.}    A few caveats about the
Peacock \& Dodds approximation are in order. It was developed to fit
simulations of power law spectra, so it can disagree significantly with N-body
results when considering power spectra that are not pure
power laws (as is the case here) or have wiggles
\cite{jenkins98,meiksin99,Jain00}. The straight mapping
between the non-linear power spectrum at one scale and the linear
power spectrum at a larger scales is only approximate, so
care should be taken when interpreting our translation from aperture
mass to $P(k)$. 

In the Peacock \& Dodds {\it Ansatz}, $k$ determines $\Delta$ which
determines $\Deltanl$, which via \eq{knlEq} determines $\knl$.  
This means
that we can think of both $\knl$ and $\P_i$ in \eq{LensingWindowDefEq}
as functions of the linear wavenumber $k$ and change variables in the
integral: 
\beq{LensingWindowDefEq2} d_i =
\int_{-\infty}^\infty\Pl_i(k)d\ln k.  
\eeq
The relation between $\knl$ and $k$ is time dependent,
so the Jacobian of the transformation cannot be taken out of the time
integral. The functions $\Pl_i(k)$ tell us which linear scales $k$ contribute to the observed
lensing signal. They are plotted in the bottom panel of \fig{lensing_windows_hoekstraFig}
for the seven data points, and their $k$-range is shown in as a function of $\theta$ in 
\fig{lensing_k_range_hoekstraFig}.

\begin{figure}[tb] 
\centerline{\epsfxsize=9.0cm\epsffile{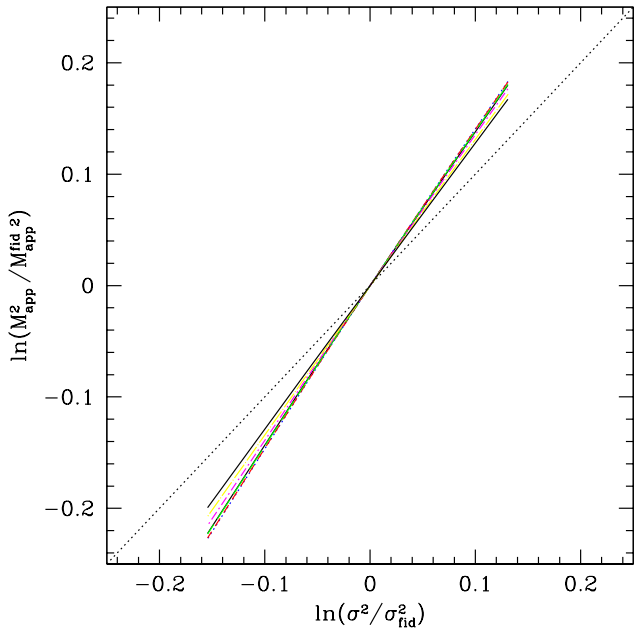}}
\smallskip
\caption{\label{lensing_gammaFig}\footnotesize%
Relation between the aperture mass variance and the overall normalization of
the matter power spectrum ($\sigma^2$). Both values are normalized to
the values for the fiducial model. Clockwise, the curves correspond to 
$\theta=$10, 15, 20, 25, 33, 42 and 50 respectively. 
Our approximation that these curves are straight lines is seen to be quite 
accurate. Their slope approaches unity 
(dotted line)  in the linear regime (for very large $\theta$).
}
\end{figure}

We see that the curves $\PPnl$ and $\Pl$ differ dramatically on small scales. Not only
do the lensing measurements probe the linear power spectrum on much larger
scales scales $k^{-1}$ than those on which it probes $\Pnl$, but the $k$-range probed
is substantially narrower as well. The relation between 
$\theta$ and (median) $k$ can be approximated  by a simple power law with half the slope over
the range of scales shown in \fig{lensing_k_range_hoekstraFig},
$k\approx (\theta/3')^{-1/2}h/\Mpc$, with the two bands converging only for
$\theta\simgt 3^\circ$ where the density fluctuations are nearly linear.
The implications of this are twofold: weak lensing probes $P(k)$ on substantially larger 
scales than a naive back-of-the-envelope calculation would suggest, and the
$k$-space window functions on $P(k)$ are quite nice and narrow, facilitating
cosmological interpretation of the measurements, although the caveats
about the Peacock \& Dodds {\it Ansatz} should be born in mind.
The last remaining subtlety involved in mapping our 
lensing measurements into $(k,P)$-space concerns the vertical placement of 
the points and their error bars. 
Since $\Pl_i(k)$ depends on $P(k)$ in a nonlinear way, we cannot simply proceed
as in the CMB case, interpreting $d_i$ as measuring a weighted average of $P(k)$.
We therefore need to construct a relation between $P(k)$ and
$d(\theta)$ around 
the fiducial model $\Pfid(k)$ (which fits the measured data from
\cite{Hoekstra02} well). 
To do so, we compute the aperture mass for models with a varying
overall normalization of $P(k)$.  As seen in \fig{lensing_gammaFig},
the relation between the aperture mass and this overall normalization
is well approximated by a straight line in log-log space whose slope
depends only on $\theta$, so we make the approximation
\beq{LinearizedLensingEq} \ln\left({P(\keff)\over\Pfid(\keff)}\right)
\approx \gamma(\theta) \ln\left({d(\theta)\over\dfid(\theta)} \right).
\eeq To translate the error bars, we simply multiply the relative
error in the aperture mass variance measurement by $\gamma(\theta)$ to
obtain the relative error in $P(k)$.

\begin{figure}[tb] 
\centerline{\epsfxsize=9.0cm\epsffile{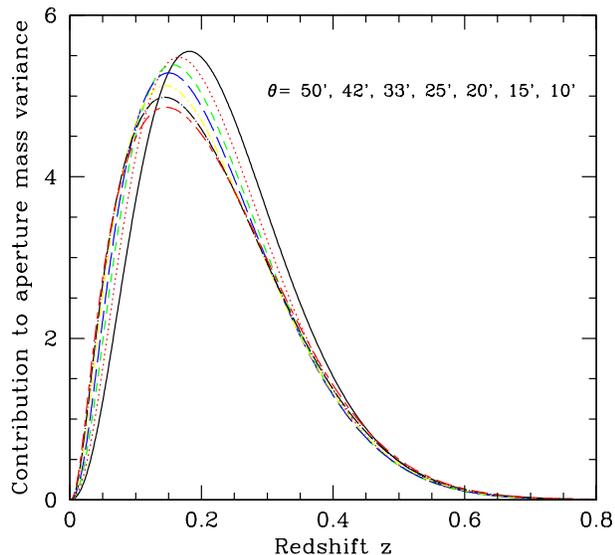}}
\smallskip
\caption{\label{lensing_zFig}\footnotesize%
Curves show $\Pz(z)$ normalized to have unit area.
These curves show where the contribution to the aperture mass
originates in redshift for each of the seven values of $\theta$
from  \protect\cite{Hoekstra02}.
}
\end{figure}

Finally, although by construction we are always mapping the
constraints 
of the different measurements onto the linear power spectrum at the
present epoch, the lensing aperture mass measurements are actually
sensitive to a weighted
average of the power spectrum over redshift, with a weight that peaks
somewhere midway between redshift zero and the redshift of the background
galaxies. Just as before we can write,  
\beq{LensingWindowredshift}
d_i = \int_{0}^\infty\Pz_i(k)dz,
\eeq 
where $z$ is redshift. For  completeness we show these integrands in figure
\ref{lensing_zFig}. These functions are seen to be very broad and to depend only weakly on
the angular scale $\theta$.

\subsection{Cluster data}

The abundance of galaxy clusters at various redshifts is emerging as
an increasingly powerful probe of cosmological parameters, as new
surveys are enlarging cluster samples and x-ray, SZ, optical and
lensing observations of cluster properties are improving our
understanding of the underlying physics.

In principle, a suite of hydrodynamical simulations including all the
relevant physics could be used to map out the region in cosmological
parameter space that matched the observed cluster abundance.  In
practice, this is still not numerically feasible, so published
constraints involve a series of approximations.  At a minimum, this
tends to involve the Press-Schechter approximation or variations
thereof \cite{PressSchechter,ShethMoTormen01,Jenkins01} to predict
the mass function of dark halos and some way of inferring the mass of
the dark halo from the observed properties of the cluster. 
For example, in studies using X-rays, a mass-temperature relation that
connects the halo mass with an observed cluster x-ray temperature is needed.
The consensus result is that cluster data constrains mainly a
combination of the normalization of the power spectrum on the cluster
scale and the cosmic density parameter $\Om$.  The normalization is
usually quoted as the rms density fluctuation $\sigma(R)$ in a sphere
of radius $R=8h^{-1}$Mpc, given by
\beq{sigmaEq} 
\sigma_R^2\equiv {4\pi\over (2\pi)^3}
\int_{-\infty}^\infty\left[{3 j_1(kR)\over(kR)}\right]^2 k^3 P(k) d\ln
k, 
\eeq
where the 1st spherical Bessel function is $j_1(x) = (\sin x-x\cos x)/x^2$.

For instance, a recent SDSS analysis reports $\sigma_8
=(0.35\pm0.03)\Om^{-0.60}$ \cite{Bahcall02}, basing cluster mass
estimates on richness rather than x-ray.
However, it has been argued \cite{Pierpaoli01} that quoting results
using $R=8h^{-1}$Mpc is confusing, since the cluster abundance is
mainly sensitive to slightly larger scales centered around $R\sim
15h^{-1}$Mpc.  The $\Om$-dependence above comes mainly from the fact
that the mean density $\Omega_m$ enters in the Press-Schechter formula
and collapse overdensity approximation, but also from a small
$\Om$-dependent correction for evolution between $z=0$ and the
redshifts observed (say $z=0.1-0.2$).  Including the additional
$\Om$-dependence coming from the fact that $\Om$ affects the shape of
the power spectrum and hence the ratio ratio $\sigma_{15}/\sigma_8$, a
result like $\sigma_8 \propto\Om^{-0.6}$ changes significantly, to
something like $\sigma_8 \propto\Om^{-0.2}$ \cite{Pierpaoli01}.

Since we wish to plot constraints on the power spectrum $P(k)$ of the form of \eq{WindowDefEq},
we follow \cite{Pierpaoli01} and use the normalization $\sigma_R$ at 
$R = 9\Omega_m^{-0.41}\approx 15 h^{-1}$Mpc. This scale corresponds to a 6.5keV cluster forming now,
and has the property of giving constraints that to first approximation depend on $P(k)$ only via its normalization  ($\sigma_R$),
not via its shape \cite{Pierpaoli01}.
Using our convention $(\star)$ to plot the $k$-range probed by clusters, 
the 20th, 50th and 80th percentiles of the integrand in \eq{sigmaEq} fall at 
$k\approx$0.06, 0.10 and 0.15$h/$Mpc, respectively.

Table 2 gives a recent sample of cluster measurements of the power spectrum normalization,
all quoted for $\Om=0.3$ for comparison.
We see that although the quoted error bars are as small as 0.05-0.08, the
spread in $\sigma_8$ between papers is many times larger, as great as $0.41$ even 
during the past year. Since this suggests that systematic uncertainties are still larger 
than statistical uncertainties, we simply use the constraint 
$\sigma_8=0.8\pm 0.2$ to be conservative, mapped to $R=15h^{-1}\Mpc$ as in \cite{Pierpaoli01} to
reduce power spectrum shape dependence.

\bigskip
\noindent
{\footnotesize
{\bf Table 2} -- Recent measurements of the $P(k)$ normalization using cluster abundances.
\bigskip
\begin{center}
{\footnotesize
\begin{tabular}{lll}
\hline
Analysis&&$\sigma_8$\\
\hline
Pierpaoli {\etal} (2001)	&\cite{Pierpaoli01}	&$1.02^{+0.07}_{-0.08}$\\
Borgani {\etal} (2001)		&\cite{Borgani01}	&$0.76^{+0.08}_{-0.05}$\\
Reiprich \& B\"ohringer (2001)	&\cite{Reiprich01}	&$0.68^{+0.08}_{-0.06}$\\
Seljak {\etal} (2001)		&\cite{Seljak01}	&$0.75\pm 0.06$\\
Viana {\etal} (2001)		&\cite{Viana01}	&$0.61\pm 0.05$\\
Bahcall {\etal} (2002)		&\cite{Bahcall02}	&$0.72\pm 0.06$\\
\hline		
\end{tabular}
}
\end{center}
}

\subsection{Ly$\alpha$ Forest data}

The Lyman $\alpha$ forest  (Ly$\alpha$F) is the plethora of absorption lines in the
spectra of distant quasars caused by neutral hydrogen in overdense
intergalactic gas along the line of sight. By tracing the cosmic gas distribution out to 
great distances, it offers a new and exciting probe of matter clustering on even smaller scales
than currently accessible to CMB and weak lensing,
when the universe was merely 10-20\% of its present age.
Since the gas probed by the Ly$\alpha$F is only overdense by a modest
factor relative to the cosmic mean, the hope is that all the relevant
physics can be simulated, thereby connecting the observations to the
underlying matter power spectrum
\cite{Narayanan00,Croft99,WhiteCroft00,McDonald00,lya}.  

The most ambitious such analysis to date \cite{Croft00} claimed to do
just this, measuring $P(k)$ on 13 separate scales $k$ using 53 quasar
spectra.  An extensive reanalysis by an independent group
\cite{GnedinHamilton01} has suggested that the technique basically
works. One should keep in mind that there are many caveats to the
Lyman $\alpha$ forest analysis. One wonders to what extent all the
relevant physics is included in the hydro-simulations and the 
dark-matter-only prescriptions that have been developed and how the
uncertainties in the reionization history, the ionizing background and
its fluctuations propagate into the reconstruction of $P(k)$. Moreover,
even for the evolution of the dark matter alone, which is the basis of
the simple {\it Ansatz} used to determine $P(k)$ from the Ly$\alpha$F data,
non-linear corrections significantly affect the evolution of clustering
on the scales relevant for the Ly$\alpha$F because the slope of the $P(k)$
around the non-linear scale is much closer to $n_{eff}=-3$ than it is today
\cite{ZaldaScoccimarroHui01}. We should view the reconstructed
points as an inversion done assuming that all the relevant physics was
correctly modeled and that the departures from the fiducial
model (which in this case also involve other details such as the
reionization history) are sufficiently small.

\Fig{pplotFig} shows the reanalyzed data \cite{GnedinHamilton01} with the quoted 
statistical and ``systematic'' errors added in quadrature. The plotted errors do not include an overall
multiplicative error of $17\%$ stemming from temperature and optical depth uncertainties,
and the mapping from the observation redshift $z\sim 2.72$ to today may introduces additional 
horizontal and vertical shifts that depend on $\Om$ and $\Ol$ as described in \sec{DegenSec}.
We use the approximation of \cite{GnedinHamilton01} that the window functions are 
Gaussian with width $\Delta k\approx 25 \ {\rm km/s}\ (k\times {\rm km/s})^{-1/2}$. 
The 13th point is a mere upper limit, omitted to avoid clutter.

\subsection{Galaxy clustering data}

Two- and three-dimensional maps of the Universe provided by
galaxy redshift surveys constitute the $P(k)$ probe with the longest tradition.
Indeed, the desire to measure $P(k)$ was one
of the prime motivations behind ever more ambitious
observational efforts such as the
the CfA/UZC \cite{Huchra90,Falco99},
LCRS \cite{Shechtman96} and PSCz \cite{Saunders00}
surveys, each well in excess of $10^4$ galaxies.
The most accurate power spectrum measurement to date is from
the 2 Degree Field Galaxy Redshift Survey (2dFGRS) \cite{Colless01},
soon to be overtaken by the 
Sloan Digital Sky Survey
(SDSS) \cite{York00} which aims for 1 million galaxies.

Band powers measured from galaxy surveys are related to the underlying matter power
spectrum by
\beq{GalaxyWindowDefEq}
d_i = \int_{-\infty}^\infty W_i(k) b(k)^2 P(k)d\ln k,
\eeq
where the window functions $W_i$ depend only on  the geometry of the
survey and the method used to analyze it. 
Here $b(k)$ is the bias, reflecting the fact that galaxies need not cluster
the same way as the underlying matter distribution, and defined 
simply as the square root of the ratio of galaxy power to matter power.
\Fig{pplotFig} shows the 2dFGRS power spectrum as measured with
the PKL eigenmode technique \cite{2df}, which has the advantage
of producing uncorrelated error bars and narrow, exactly computable window functions $W_i$
(see also \cite{Percival01}).

With galaxy clustering measurements, bias is the key caveat.
On small-scales, bias is known to be complicated, with the galaxy power spectrum
saying more about the galaxy distribution within individual dark matter halos than about the
underlying matter distribution. We have therefore plotted 2dFGRS measurements only for
$k<0.3h/\Mpc$.
Fortunately, a broad class of bias models predict that $b(k)$ should be simple and independent of $k$ 
on large scales 
\cite{Coles93,Fry93,Scherrer98,Coles99,HeavensMatarreseVerde99}.
Even if this is true, however, the measured large-scale 2dFGRS power spectrum is likely to have slightly
scale-dependent bias, masquerading as evidence
for a redder power spectrum, \ie, one with a smaller
spectral index $n$.
This is because the power spectrum is measured from a 
heterogeneous magnitude-limited sample mixing galaxies of very different kinds.
Most of the information about $P(k)$ on large scales comes
from distant parts of the survey, where bright ellipticals
are over-represented since dimmer galaxies get excluded by
the faint magnitude limit. 
Since more luminous galaxies are known to be more highly biased
\cite{Zehavi02,Norberg01}, this should cause 
the bias to rise as $k\to 0$.
With a massive data set like the SDSS, it will be possible to accurately measure how bias depends 
on luminosity and correct for this effect.

\section{Breaking the degeneracy between primordial power and transfer functions}
\label{DegenSec}

Above we have shown how to map CMB, lensing, cluster and Ly$\alpha$F measurements into $k$-space when the transfer functions are known.
These transfer functions depend on various Late Universe cosmological
parameters (the reionization optical depth $\tau$ and the matter budget). 
To measure the Late Universe properties (these parameters) and the Early Universe properties
(the primordial power spectrum $\Pstar(k)$) independently, we must therefore 
break the degeneracy between the two. 
This may at first sight appear hopeless, since the measurements involve products
of primordial power and transfer functions, and there is no unique way of factoring 
a product into two terms. As will be described below, the problem can 
nonetheless be solved thanks to two separate facts in combination:
\begin{enumerate}
\item $\tau$ and the matter budget parameters affect different types of measurements in different ways.
\item There is substantial overlap in $k$-space between different types of measurements.
\end{enumerate}

\begin{figure}[tb] 
\centerline{\epsfxsize=9.0cm\epsffile{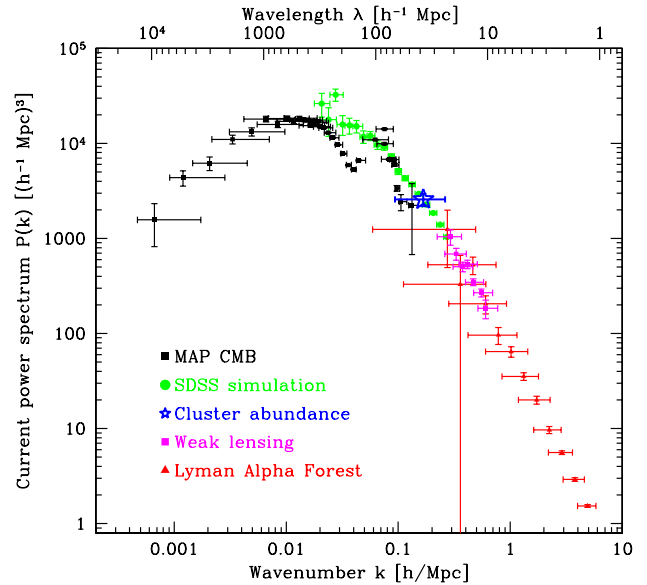}}
\vskip-0.8cm
\smallskip
\caption{\label{pplot_wrong_obFig}\footnotesize%
Same as \fig{pplot_futureFig}, but assuming a baryon density $h^2\Omega_b=0.07$
when analyzing the data. 
Since this figure makes no assumptions about the primordial power spectrum $\Pstar(k)$,
the glaring discrepancy seen between the power spectrum inferred from CMB and 
galaxy clustering means that this high baryon fraction can be ruled out without assumptions
about Early Universe physics.
}
\end{figure}

A picture can say more than a thousand words, and both of these facts are illustrated 
by the example in \fig{pplot_wrong_obFig}.
It shows simulated data assuming that the concordance model of \cite{Efstathiou02} is true,
with the CMB mapped into $k$-space assuming a higher baryon density, 
$h^2\Omega_b=0.07$. This alters the CMB and matter transfer functions
in quite different ways (details below), producing a strikingly wiggly $P(k)$ inferred from the CMB.
Since MAP and SDSS overlap by over a decade in $k$ where this wiggliness is seen, 
it is obvious that the two are inconsistent and that such a high baryon density is ruled out.
This conclusion can be drawn without assumptions about the primordial power spectrum $\Pstar(k)$,
since this figure was generated without involving $\Pstar(k)$, merely 
using measurements and transfer function parameters.

In the next subsection, we will briefly discuss how the various types of measurements are
affected by the Late Universe parameters and the relevance of this for measuring these
parameters independently of $\Pstar(k)$. We then describe how how a ``chi-by-eye'' comparison
as in \fig{pplot_wrong_obFig} can be replaced by a rigorous statistical method useful for
cosmological parameter estimation.

\subsection{How Late Universe parameters affect the $P(k)$ recovered from different data sets}
\label{ParameterDependenceSec}

Although reconstruction of $P(k)$ as in \fig{pplot_wrong_obFig} has the advantage of 
minimizing the amount of processing applied to large-scale-structure 
(galaxy, lensing, cluster, Ly$\alpha$F) data
the reconstructed primordial power $\Pstar(k)$ provides better intuition for the present discussion,
since each data set has, loosely speaking, been divided by its own transfer function. 
In contrast, the CMB data in \fig{pplot_wrong_obFig} was 
(apart from smoothing effects)
both divided by the CMB transfer function $W_\l(k)$ and multiplied by the matter transfer 
function $T(k)^2$. We will therefore center our discussion around  $\Pstar(k)$ rather than $P(k)$
in the remainder of the paper.

\begin{figure}[tb] 
\centerline{\epsfxsize=9.0cm\epsffile{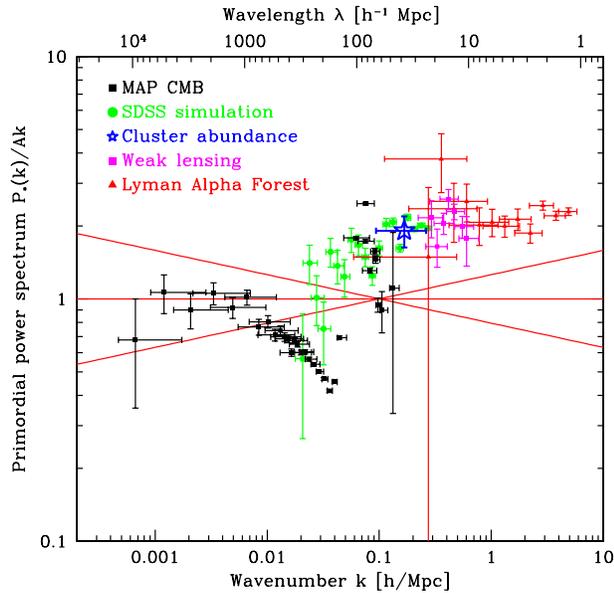}}
\vskip-0.8cm
\smallskip
\caption{\label{kplot_wrong_obFig}\footnotesize%
Same as \fig{kplot_futureFig}, but assuming a baryon density $h^2\Omega_b=0.07$
--- a ruled out model. 
}
\end{figure}

\begin{figure}[tb] 
\centerline{\epsfxsize=9.0cm\epsffile{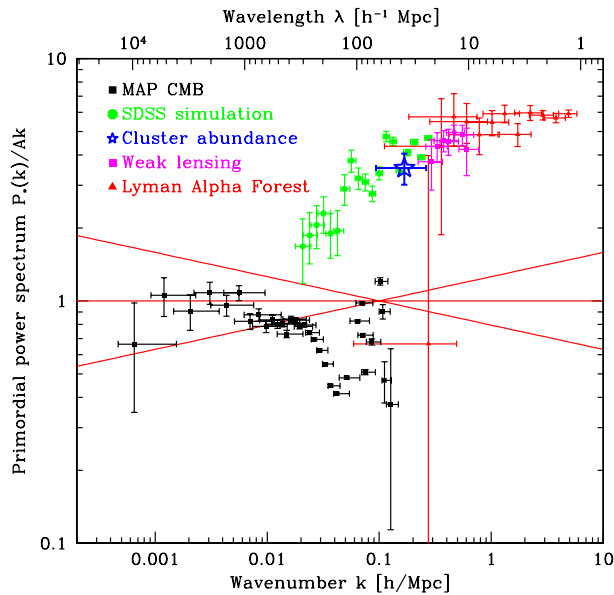}}
\vskip-0.8cm
\smallskip
\caption{\label{kplot_wrong_ocFig}\footnotesize%
Same as \fig{kplot_futureFig}, but assuming a dark matter density $h^2\Omega_c=0.5$ 
--- a ruled out model. 
}
\end{figure}

\begin{figure}[tb] 
\centerline{\epsfxsize=9.0cm\epsffile{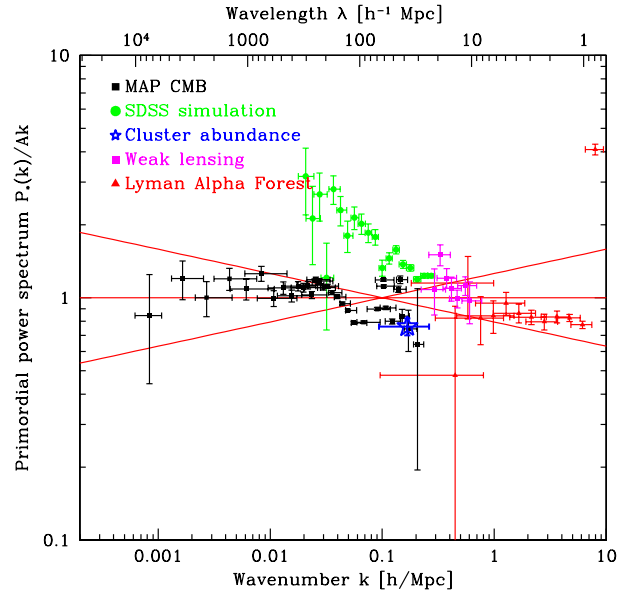}}
\vskip-0.8cm
\smallskip
\caption{\label{kplot_wrong_OlFig}\footnotesize%
Same as \fig{kplot_futureFig}, but assuming a cosmological constant $\Omega_\Lambda=0.2$
--- a ruled out model. 
}
\end{figure}

Imagine generating large numbers of plots like \fig{kplotFig}, each one assuming
different values for the Late Universe parameters to analyze the same measured data.
\Fig{kplot_wrong_obFig} shows the result for the above-mentioned case of a high
baryon fraction, and is simply the $\Pstar$-version of \fig{pplot_wrong_obFig}.
Figures~\ref{kplot_wrong_ocFig} and \ref{kplot_wrong_OlFig} show corresponding examples
with incorrect assumptions for the cold dark matter density $h^2\Oc$ and the
cosmological constant $\Ol$.
Before delving into details, a few basic facts should be noted. 
The LSS (cluster, lensing, Ly$\alpha$F and galaxy) points tend to shift together, since they are
all sensitive to the matter transfer function $T(k)^2$. These LSS points split apart when certain parameters
are altered, however, notably $\Om$ and the cosmic expansion history $a(t)$, which affect
the four differently. In contrast, the CMB points separate from the other four data types whenever
any Late Universe parameter is changed, indeed often by shifting in a rather opposite
direction from the others.

The recent literature on cosmological model constraints includes a bewilderingly large
list of cosmological parameters:
\beq{pEq}
\p\equiv(\tau,\Ok,\Ol,\Ox,w,\od,\ob,\fn,\As,\ns,\alpha,\At,\nt).
\eeq
These are the reionization optical depth $\tau$,
the primordial amplitudes $\As$, $\At$ and tilts $\ns$, $\nt$
of scalar and tensor fluctuations, the running of the scalar tilt $\alpha$,
and seven parameters specifying the cosmic matter budget.
The various contributions $\Omega_i$ to critical density are for
curvature $\Ok$, vacuum energy $\Ol$, other dark energy $\Ox$ (with an equation of state $w$), 
cold dark matter $\Oc$,
hot dark matter (neutrinos) $\On$ and baryons $\Ob$.
The quantities
$\ob\equiv h^2\Ob$ and
$\ocdm\equiv h^2\Od$ correspond to
the physical densities of baryons
and total (cold + hot) dark matter
($\Od\equiv\Oc+\On$), and $\fn\equiv\On/\Od$ is the fraction
of the dark matter that is hot.
Additional parameters that are often mentioned are not independent,
for instance the total matter density $\Om\equiv\Ob+\Od$ and
the dimensionless Hubble parameter
$h\equiv\sqrt{(\ocdm+\ob)/(1-\Ol\Ok)}$.

Fortunately, the underlying physics is simpler than this parameter
profusion suggests.  $\As$, $\ns$ and $\alpha$ are merely a particular
parametrization of the primordial power spectrum, corresponding to the
{\it {\it Ansatz}} $\Pstar(k)=\As k^{\ns +\alpha\ln k}$.  $\At$ and
$\nt$ similarly parametrize the primordial tensor (gravity wave) power
spectrum as a power law.  In other words, only the first eight
parameters in \eq{pEq} are Late Universe parameters affecting the
transfer functions --- we refer to all of these except $\tau$ as the
matter budget parameters.  The tensor parameters $\At$ and $\nt$ are
of only marginal relevance to this paper, since they affect only the
CMB and do so essentially only on scales larger than those that
overlap with large-scale-structure observations.  This means that if
we assume $\At>0$ in our reconstruction, the CMB measurements of
$\Pstar(k)$ would shift downwards in \fig{kplot_wrong_obFig}, but only
to the left where they cannot be compared with other data. Since our
method for measuring Late Universe parameters involves comparing CMB
with LSS data, it is therefore essentially unaffected by tensor
fluctuations.

A similar simplification applies to $\tau$, which also affects only the CMB.
 On the small scales where the CMB overlaps with other 
measurements, the effect of reionization in merely to suppress the CMB power spectrum by 
a constant factor $e^{2\tau}$. 

A further simplification is that $\Ol$, $\Ox$ and $w$ never enter in
any other way than as a particular parametrization of the cosmic
expansion history $a(t)$ or, equivalently, of the function
\beq{Heq}
{H(z)\over H_0} = \left[(1+z)^3\Om + (1+z)^2\Ok +  (1+z)^{3(1+w)} + \Ol\right],
\eeq
where the Hubble parameter $H\equiv d\ln a/dt$\footnote{If the dark
energy is a scalar field that can cluster (ie. quintessence) there
could be additional effects for low $l$s.}. Various integrals involving this function
determine the growth of linear clustering, 
the brightness and angular size of distant objects, and volume-related effects.
This function causes merely a sideways shift in the CMB on the scales that LSS can probe,
since the late ISW effect is important only on larger scales.
Beyond the parameters in this function, the only remaining matter budget parameters
are thus $\om$, $\ob$, $\fn$, specifying the physical densities of cold dark matter, baryons and 
neutrinos, respectively.

Detailed discussions how the CMB and matter transfer functions depend on cosmological 
parameters can be found in, \eg, \cite{HuNature97,EisensteinHu99,Kosowsky02}. 
For the reader interested in 
more empirical intuition, the movies at $www.hep.upenn.edu/{\sim}max/cmb/movies.html$
are recommended.
The key point about to take away from all this is that the CMB and matter transfer functions
depend quite differently on the matter budget parameters, often in rather opposite ways.
For instance, increasing the cold dark matter density $\ocdm$ shifts the galaxy power
spectrum up to the right and the CMB peaks down to the left.
Adding more baryons boosts the odd-numbered CMB peaks
but suppresses the galaxy power spectrum rightward of its peak and also makes it wigglier. 
Increasing the dark matter percentage that is hot (neutrinos) 
suppresses small-scale galaxy power while leaving the CMB almost unchanged.
(The recovered points in our $\Pstar(k)$ figures 
always respond in the sense opposite to that in these movies when assumed parameters are varied.)
This means that combining CMB with LSS data allows unambiguous determination
of the matter budget.

Since the heights of the CMB peaks are controlled by the 
densities of ordinary ($\ob$) and dark ($\od$) matter, assuming the wrong values for these
parameters is seen to results in a wiggly $\Pstar(k)$ from CMB in 
Figures~\ref{kplot_wrong_obFig} and \ref{kplot_wrong_ocFig}.
Increasing $\ob$ boosts predominantly the odd-numbered CMB peaks whereas 
increasing $\ocdm$ suppresses the CMB peaks with less of an even/odd asymmetry
as well as shifting the peaks to the left, so the CMB points in these figures are seen to depart from
unity in the opposite sense. 
Increasing the baryon fraction $\ob/\od$ produces larger wiggles in the matter power
spectrum as well, together with an overall power suppression leftward of the peak.
Increasing the dark matter density $\od$ pushes the turnover in $P(k)$, corresponding
to the horizon size at the matter-radiation equality epoch, to the right and thereby 
boosts small-scale power. 
There is also a sideways shift in both CMB and matter clustering, since the $h$ on the 
horizontal axis changes with the matter budget parameters.
$H(z)$ gives shifts the transfer functions vertically
via the linear growth factor and horizontally
(angle-diameter distance changes, and well as the $h$ in the horizontal axis definition).
Here is a brief summary of how the curves $\Pstar(k)$ recovered from CMB and LSS measurements
get affected when the assumed parameters are changed:
\begin{itemize}
\item $\ob$, $\ocdm$: cause wiggles
\item $\fn$: boosts LSS points on small scales
\item $\Ok$, dark energy: cause wiggles via incorrect CMB peak locations, vertical offset
\item $\tau$: boosts CMB points by $e^{2\tau}$ where they overlap with LSS
\end{itemize}
Some parameters also affect the conversion of observed LSS data to measurements of $P(k)$
as illustrated in \fig{kplot_wrong_OlFig}.
The cluster point scales roughly as $\Om^{-1.2}$ in power 
($\Om^{-0.6}$ in amplitude), mainly because $\Om$ enters in the Press-Schechter prescription
for halo abundance and in the approximation for collapse overdensity. 
For weak lensing, a factor of $\Om^2$ enters in the definition of 
the cosmic shear power spectrum, although the final scaling with $\Om$ is complicated by nonlinearities.
All four LSS observables at $z>0$ must be mapped to $z=0$, which involves a 
vertical shift due to clustering growth and a horizontal shift from the computation of 
comoving length scales, and these shifts are both given by the cosmic expansion history
$H(z)$ of \eq{Heq}. Since these shifts increase with $z$, they are most important for the
Ly$\alpha$F.
Finally, the $P(k)$ reconstruction from galaxies, clusters and the Ly$\alpha$F
depend at least weakly on the baryon fraction $\Ob/\Om$, as is evident from considering 
what entities like galaxy bias, the cluster temperature function would be like in the limit
of zero baryons.

\begin{figure}[tb] 
\centerline{\epsfxsize=9.0cm\epsffile{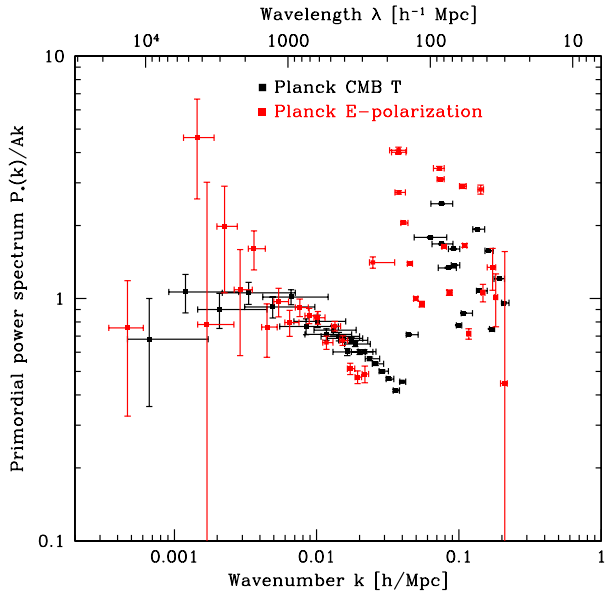}}
\vskip-0.8cm
\smallskip
\caption{\label{kplot_planckFig}\footnotesize%
Simulation of the primordial power spectrum recovered from the upcoming 
Planck CMB satelle using unpolarized and polarized data,  assuming a baryon density $h^2\Omega_b=0.07$
--- a ruled out model. Note that the bias goes in opposite directions, since polarized and
unpolarized peaks are out of phase. The simulation was performed using all Planck channels and 
the MID foreground model from \protect\cite{foregpars}.
}
\end{figure}

CMB polarization provides another powerful and independent tool for breaking the
degeneracy between Early and Late Universe parameters. The key reason for this is that
the polarized acoustic peaks are out of phase with the unpolarized ones, with the peaks in the
E power spectrum lining up with troughs in the unpolarized (T) power spectrum. 
This means that an incorrect assumed value for a parameter affecting peak heights 
(notably $\ob$ and $\ocdm$) causes the primordial power spectrum $P_*(k)$ recovered 
from T and E will to be biased in opposite directions. This useful fact is illustrated in 
\fig{kplot_planckFig}, which shows simulated measurements from the Planck satellite
mapped into $k$-space. Note in particular the wiggle at $k\sim 0.04h/\Mpc$, where
the T-points go low whereas the E-points go high, revealing that the high baryon 
fraction assumed is ruled out at high significance from CMB data alone, without any
assumptions about the shape of the primordial power spectrum.

\def\pstark{\Pstar(k)}  
\subsection{Measuring Late Universe parameters independently of $\pstark$}

Above we described the physics that makes it possible to break the
degeneracy between Early Universe parameters ($\Pstar(k)$)
and Late Universe parameters ($\tau$ and the matter budget). We now turn to the issue
of how to do this in practice in a statistically rigorous way.

\subsection{The basic problem}

Our basic problem is to measure a vector $\p$ of  Late Universe parameters, say
\beq{p2Eq}
\p\equiv(\tau,\od,\ob,\fn,\Ok,\Ol),
\eeq
independently of $P_*(k)$. Our basic approach is to map all measurements into
(linear) $k$-space and test if they are consistent with one another. Repeating this
for a fine grid of models $\p$, we can map out the region of parameter space that is 
allowed, \ie, where the data are consistent. 

To be specific, let us consider the case of comparing two types of data,
for instance the MAP CMB power spectrum with the SDSS galaxy power spectrum.
The challenge is that although they exhibit substantial overlap in scale, as seen in 
\fig{kplot_futureFig}, they generally have different window functions. This means that we cannot simply 
subtract the two independent measurements of $\Pstar(k)$ from each other and 
require the result to vanish --- it would not vanish even in the absence of measurement
errors, since the two experiments are probing different linear combinations of the underlying
function $\Pstar(k)$.

Fortunately, a mathematically equivalent problem has already been extensively investigated in 
the literature, and we can employ the solution to test our data sets for consistency.
The question of whether there is any power spectrum that is consistent with 
given band power measurements was studied in \cite{Juszkiewicz87} for the case of two 
band powers and generalized to multiple ones in \cite{pindep}. For $n$ band-powers, 
the best fit power spectrum was found to be a linear combination of up to 
$n-1$ delta-functions \cite{pindep}. However, finding their locations is 
numerically time-consuming for large $n$, which motivated revisiting the problem.
A fast general method was presented in \cite{comparing} but was somewhat complicated, 
involving a series of eigenvalues computations.
Finally, an information-theoretically optimal method was derived in \cite{qmaskpow}
for different purposes (comparing CMB maps), which solves our present problem as well.
This method has also been used to test CMB experiments for consistency
\cite{qmaskpow,consistent}.
Since it is not only better (in terms of statistical rejection power) 
but also simpler than the above-mentioned alternatives, we will summarize it briefly below.
It consists of two parts: ``deconvolving'' the effect of the window functions and 
testing the resulting two measurements for consistency.
In summary, our method thus consists of  the following three steps:
\begin{enumerate}
\item Map the measurements into $k$-space for fixed a $\p$
\item Deconvolve the effect of window functions
\item Test the deconvolved measurements for consistency
\end{enumerate}
These three steps are repeated for a fine grid of $\p$-vectors to map out the
allowed region in $\p$-space.
The implementation of Step 1 was described in \sec{Psec} and we now turn to 
Step 2 and Step 3.

\subsubsection{Deconvolving the effect of window functions}

Let us model the primordial power spectrum $\Pstar(k)$ as piecewise constant in
a large number of narrow $k$-bins, with $x_i$ denoting the value of $\Pstar$ in the
$\ith$ bin, and arrange the numbers $x_i$ into a vector $\x$.
Grouping our two sets of band-power measurements 
into vectors $\y_1$ and $\y_2$, this means that they are related to $\x$ by 
\beq{yModelEq}
\y_1=\W_1\x+\n_1,\quad \y_2=\W_2\x+\n_2,
\eeq
for some known window matrices $\W$ that are simply discretized versions of the
window functions computed in \sec{Psec} and some noise vectors with known statistical 
properties. We assume that the measurements are unbiased so that 
$\expec{\n_i}=\bzero$ and define the noise covariance matrices
\beq{NdefEq}
\N_1\equiv\expec{\n_1\n_1^t},\quad \N_2\equiv\expec{\n_2\n_2^t},
\eeq
In this subsection, we describe how the annoying $\W$-matrices can be eliminated by computing 
two deconvolved data sets $\x_i$, $i=1,2$,  with the properties
\beq{xiEq}
\x_i=\x+\n'_i,\quad \expec{\n'_i}=\bzero,
\eeq
and known covariance matrices 
$\NN_i\equiv\expec{{\n'_i}{\n'_i}^t}$.

In the generic case, such deconvolution is strictly speaking impossible:
we cannot compute $\W_i^{-1}\y_i$ since $\W_i$ is not invertible. 
Certain pieces of information about $\x$ are simply not present in $\y_i$,
for instance about sharp features on scales much smaller than the widths of the window
functions or about $k$-scales outside the region probed by the observations.
The basic idea in Appendix D of \cite{qmask} is to 
accept that certain modes in $\x$ cannot
be recovered, and to record this information in the noise covariance
matrix $\NN_i$ for $\x_i$ by assigning a huge variance to these modes.
Any subsequent analysis (in our case consistency testing)
will then automatically assign essentially zero weight to these modes.
This is useful in practice since all complications related to 
window functions are transferred from $\W_i$ to $\NN_i$ where, as we will see
in the next subsection, they are straightforward to deal with. 

The method can be interpreted as 
combining the real data $\y_i$ with data from a virtual experiment
that is so noisy that it contains
essentially no information, yet has enough information to remove all numerical 
singularities by providing independent measurements of each $x_i$ with
some huge standard deviation $\sigma$. 
The final result is \cite{qmask}
\beqa{newsimEq}
\x_i&=&\NN_i\W_i^t\N_i^{-1}\y_i,\\
\label{newsimSig}
\NN_i&=&\left[\W_i^t\N_i^{-1}\W_i + \sigma ^{-2}\I\right]^{-1}.
\eeqa
One finds \cite{qmask} that this prescription has all desired properties
as long as $\sigma$ is chosen to be a few orders of magnitude larger than the error bars 
in the real data, and it also has the property of minimizing the noise variance 
in the deconvolved data.
If we were to choose $\sigma$ to be too small, then the virtual experiment
would contribute a non-negligible amount of information and bias the
results. If we were to choose $\sigma$ to be too large, however,
the matrix $\NN_i$ would contain some enormous eigenvalues
(since $\W_i^t\NN_i^{-1}\W_i$ is typically not invertible)
and be poorly conditioned, causing numerical problems.

\subsubsection{Testing the deconvolved measurements for consistency}

To make explicit that our mapping of measurements into $k$-space depends on the
assumed Late Universe parameters $\p$, we will denote the 
deconvolved measurements and their noise covariance matrices
$\x_i(\p)$ and $\NN_i(\p)$ from now on $(i=1,2)$.
Given these two deconvolved measurements $\x_1(\p)$ and $\x_2(\p)$ of the 
primordial power spectrum vector $\x$ (from MAP and SDSS, say), 
we wish to test if they are
consistent. If they are not, this rules out the Late Universe parameters $\p$
that were assumed to construct them.

Letting $\z$ denote the difference of the two power spectrum measurements,
\beq{zDefEq}
\z(\p)\equiv\x_1(\p)-\x_2(\p),
\eeq
we consider two hypotheses: 
\begin{itemize}
\item[$H_0$:]The null hypothesis that the assumed 
Late Universe parameters $\p$ are correct. Then $\expec{\x_i(\p)}=\x$,  
so that the difference spectrum $\z(\p)$ consists of pure noise
with zero mean and covariance matrix 
$\NN(\p)\equiv\expec{\z(\p)\z(\p)^t}=\NN_1(\p)+\NN_2(\p)$.
\item[$H_1$:]The alternative hypothesis that the assumed 
Late Universe parameter vector $\p$ is incorrect. In this case, 
$\z(\p)$ is expected to on average 
depart more from zero than under the hypothesis $H_0$, since both 
$\x_1(\p)$ and $\x_2$(\p) become
biased power spectrum measurements and typically get biased in quite different ways
(Figures~\ref{kplot_wrong_obFig}-\ref{kplot_wrong_OlFig}).
\end{itemize}

We will model $H_1$ as a change in the mean,
$\expec{\z}=\m\ne\bzero$. It can be shown \cite{comparing,consistent}
that the ``null-buster'' statistic 
\beq{NullbusterEq}
\nu(\p)\equiv\frac{\z(\p)^t\NN(\p)^{-1}\Q\NN(\p)^{-1}\z(\p) -
\tr\{\NN(\p)^{-1}\Q\}}
{\left[2\>\tr\{\NN(\p)^{-1}\Q\NN(\p)^{-1}\Q\}\right]^{1/2}} 
\eeq 
rules out the null hypothesis $H_0$ with the largest average significance
$\expec{\nu}$ if $H_1$ is true provided one chooses $\Q=\m\m^t$. The
statistic can be interpreted as the number
of ``sigmas'' at which $H_0$ is ruled out \cite{comparing}.  
Note that for the special case
$\Q\propto\NN(\p)$, it simply reduces to $\nu=(\chi^2-n)/\sqrt{2n}$,
where $\chi^2\equiv\z^t\NN(\p)^{-1}\z$ is the standard chi-squared
statistic.  The null-buster test can therefore be viewed as a
generalized $\chi^2$-test which places more weight on those particular
modes where the expected signal-to-noise is high.  It has proven
successful comparing both CMB maps \cite{qmask,qmap3,qmaskpow}, galaxy
distributions \cite{r,EfstathiouNullbuster} and CMB power spectra
\cite{consistent}.  Tips for rapid implementation in practice are
given in \cite{qmask}.

Any choice of $\Q$ results in a statistically valid test: for instance, the
region in the Late Universe parameter space for which
$\nu(\p)>2$ is ruled out at the $2-\sigma$ level. This corresponds to 97.5\% 
significance with the usual frequentist interpretation as a one-sided test if there are 
many similarly large eigenvalues of 
$\NN^{-1/2}\Q\NN^{-1/2}$, since it makes the generalized
$\chi^2$ distribution of $\nu$ approximately Gaussian by the 
central limit theorem.

What is the best choice of $\Q$?  The above-mentioned choice
$\Q=\m\m^t$ placed all weight on a single mode $\m$.  More generally,
the test pays the greatest attention to those eigenvectors of $\Q$
whose eigenvalues are large.  The standard $\chi^2$-test (the choice
$\Q=\NN$) has the attractive feature of giving an assumption-free
consistency test: since $\Q$ has no vanishing eigenvalues, it is
sensitive to any type of discrepancies between the two power spectrum
measurements $\x_1(\p)$ and $\x_2(\p)$. 

If the goal is to place sharp
constraints on cosmological parameters, however, the statistical
rejection power of the test can be boosted by placing the statistical
weight on precisely those types of departures which correspond to
incorrect parameter assumptions. If the rms uncertainties on the
parameters $p_i$ can be estimated using the Fisher matrix \cite{karhunen}
$\F_{ij}={\partial \m^t/\partial p_i}\NN^{-1}{\partial\m/\partial p_j}$
(with $i$ and $j$ running over the parameters),
the choice
\beq{OptimalQeq}
\Q\equiv
\sum_{ij}
(\F^{-1})_{ij} {\partial\m \over\partial p_i}{\partial \m^t\over\partial p_j}
\eeq
will have this attractive property as long as the $\p$-vectors considered
are fairly close to the true value --- a desirable situation that the cosmology community 
will hopefully keep approaching as data keeps improving.

Indeed, in this high signal-to-noise limit, the nullbuster test with the weighting given by
\eq{OptimalQeq} can be thought of as the frequentist analog of a Bayesian likelihood
analysis. To understand this we can calculate the expectation value of
$\nu$ assuming $H_1$ is true to lowest order in the parameter
differences $\Delta p_i$. To do so we take the expectation value of
equation (\ref{NullbusterEq}), use $\expec{\z\z^t}=\NN+\m\m^t$ and
$\m \approx ({\partial \m^t/\partial p_i})\Delta p_i$ and get,
\beq{meannu}
\expec{\nu}\approx{\Delta {\bf p}^t \F {\Delta \bf p}\over \sqrt{2 N_p}},
\eeq
where $N_p$ is the number of parameters. We can compare this result to
what would be obtained using a likelihood analysis. In the Gaussian
approximation, the likelihood of
the null hypothesis is given by
\beq{Leq}
\L \propto |\NN|^{-1/2}e^{-{1\over 2} \z^t\NN^{-1}\z}.
\eeq
Thus if we calculate the expected value of the logarithm of this
likelihood assuming again that $H_1$ is true (again to lowest order 
in $\Delta{\bf p}$) we get 
\beq{expectlike}
\expec{\ln \L} \approx -{1\over 2} {\Delta \bf p}^t \F {\Delta \bf p}
\eeq
up to an irrelevant additive constant,
the usual result that the Fisher matrix gives the curvature of the
log-likelihood around the maximum. This shows that 
on average and to lowest order in $\Delta\p$, the (frequentist) 
nullbuster statistic  coincides with the log-likelihood that one
would calculate in a Baysian framework, and so on average one would rule
out the same region of parameter space with both methods.

Note that although it is tempting to perform a standard Bayesian likelihood analysis using
a  ``likelihood'' similar to \eq{Leq}, it is not obvious that this can be given the
standard interpretation in terms of Bayes' theorem, since it
is not only $\expec{\z}$ and $\NN$ that depend on $\p$ (as usual), but
also the data itself: $\z=\z(\p)$. Since there are no such interpretational issues 
with the frequentist null-buster test, it is therefore reassuring that 
the two approaches approximately agree.

\subsection{Further applications}

Above we described how requiring consistency between two sets of
measurements (say MAP and SDSS) could constrain the Late Universe
parameters without assumptions about the primordial power spectrum
$\Pstar(k)$.  Given three or more types of data, an obvious extension
is to first test two for consistency (say lensing and Ly$\alpha$F),
then for the parameter subspace where they are consistent, combine
them as in \cite{comparing} and test the result against a third type
of data. Since the LSS data share many parameter dependencies as
discussed in \sec{ParameterDependenceSec}, it is natural to first let
galaxy clustering, lensing, cluster and Ly$\alpha$F measurements slug
it out amongst themselves and then compare the resulting spectrum
combining all LSS data with CMB.

Once $\p$ has been constrained as described above, it becomes possible to 
measure $\Pstar(k)$ without assumptions about $\p$ for the very first time.
One way to do this is to measure $\Pstar(k)$ as in 
\fig{kplot_futureFig} using the best fit $\p$-vector, and then quantify the error bars and their
correlations by recomputing  $\Pstar(k)$ for a grid of models $\p$ in the 
allowed region of the Late Universe parameter space, effectively marginalizing over
$\p$ using the observed constraints as a prior.

\section{Discussion}
\label{DiscSec}

We have presented a method which complements the traditional ``black box'' 
likelihood approach to cosmological parameter estimation in two ways:
by testing underlying physical assumptions and by improving physical
intuition for where the constraints come from.
We described how CMB, galaxy, lensing, cluster and Ly$\alpha$F could
be compared directly in (linear) $k$-space in \sec{Psec}, then showed how a
graphical chi-by-eye test could be transformed into a statistically 
rigorous method in \sec{DegenSec}, providing independent measurements
of Early Universe parameters (the primordial power spectrum $\Pstar(k)$)
and Late Universe parameters ($\tau$ and the matter budget).
We found that requiring consistency between unpolarized CMB measurements and either polarized
CMB or large-scale structure data  is quite promising in this regard.

Separating Early and Late Universe physics is particularly timely given the 
excess in the small-scale CMB power spectrum recently reported by the CBI  team \cite{Mason02}.
We have seen that the angular scales where this excess is seen correspond to spatial scales 
around $k\sim 0.2h/\Mpc$ where the power spectrum is already constrained by 
galaxy, lensing and cluster observations. This makes it difficult to blame the excess on the
Early Universe, say  by tilting or adding a feature in $\Pstar(k)$. The alternative explanation in
terms of contamination from discrete SZ-sources has been shown to be extremely
sensitive to the power spectrum normalization on the cluster scale, 
tentatively requiring $\sigma_8\approx 1$ \cite{Bond02}.
With our convention $(\star)$, $\sigma_8^2$ probes the range
$k\approx 0.17^{+0.09}_{-0.08}$, which according to \fig{keffFig} corresponds approximately
to the multipole range $\l=2000\pm 800$ 
if the concordance model \cite{Efstathiou02} is not too far from the truth.
The cluster normalization $\sigma_R^2$ with $R\approx 15h^{-1}\Mpc$ corresponds 
to $\l\approx 1100^{+600}_{-500}$, scales already probed by the shallower 
(mosaic) CBI observations \cite{CBI}.
The concordance model \cite{Efstathiou02} normalized to the CMB data gives
$\sigma_8=0.81$ assuming $\tau=0.05$, and it appears likely that quite accurate and robust
$\sigma_8$-measurements based on the CMB normalization will be available down the road.

This paper is not intended to the final word on testing physical assumptions in cosmology,
merely a small step to be followed by many more.
In this spirit, let us close by summarizing some of the most important things that 
we have not done and some promising directions for future work.

An obvious first step is implementing our method to independently measure
the Early and Late Universe parameters. 
Next year will be an appropriate time to do this, taking advantage of the
revolutionary precision that will be offered by MAP. 
Since this will involve working with a large grid of CMB transfer 
functions, the approximations described in \cite{DASH} will be useful for this.

We have made one important assumption throughout this paper that it would
desirable to test: that the primordial fluctuations are adiabatic. 
The adiabatic assumption means that the process generating the fluctuations in 
the Early Universe created density fluctuations without altering the density ratios of
different matter components (photons, baryons, neutrinos, dark matter, \etc). 
By tinkering with these relative densities, it is possible to generate a variety of 
so-called isocurvature fluctuations. 
In addition to the familiar baryonic and CDM isocurvature modes, there are
obscure ones, for instance a neutrino isocurvature mode where 
the ratio of neutrinos to photons varies spatially but the net density perturbation vanishes,
and it can be shown that the most general case corresponds to a function $\Pstar(k)$ that is not a scalar but
a $5\times 5$ symmetric matrix \cite{Bucher00}.
Although it has been shown that CMB polarization will help constrain isocurvature modes, 
real progress in this endeavor is likely to be some way off, 
requiring the sensitivity of the Planck satellite \cite{Enqvist00,Bucher01,Amendola02}.
On the bright side, these complications matter only when studying CMB data, so it is valid 
to compare and combine the power spectra measured with galaxies, Ly$\alpha$F, lensing and clusters
at low ($z\simlt 10$, say) redshift as described above assuming purely adiabatic fluctuations.

Staying on the topic of still more general tests, there exist an elegant 
``generalized dark matter'' formalism for describing the gravitational
effects of the most general matter budget \cite{HuGDM}, and this is likely to place
robust constraints on dark matter and dark energy as power spectrum measurements
continue to accumulate over a range of redshifts.

Needless to say, CMB and LSS data will keep improving dramatically in coming years,
providing greater sensitivity, $z$-range, $k$-range and $k$-resolution.
In addition to the obvious advantages of sensitivity and range, 
galaxy window functions will become narrower as the SDSS sky coverage improves, 
improving the ability to detect sharp features in $\Pstar(k)$.
For lensing, the shear power spectrum contribution from a given $z$ has
delta function windows on $P(k)$ in the small-angle approximation,
so the smearing in $k$-space comes from projection effects and 
can probably be substantially reduced using photometric redshift and topography techniques.
For the Ly$\alpha$F, the $\sim 10^5$ SDSS quasar redshifts should greatly improve the 
measurement errors on larger spatial scales, where one may hope that
uncertainties associated with nonlinear physics are smaller.
In addition to the CMB and LSS probes we have utilized in this paper, many more appear promising.
For instance, the recently claimed detection of galaxy halo substructure \cite{Dalal02} falls 
right on our concordance curve in \fig{pplotFig} but two orders of magnitude further right than
the other data points, at $k\sim 100\Mpc/h$. 
Searches for phase space clumpiness in the Milky Way halo with
tidal streamers and future space-based astrometry may provide further constraints on 
dark matter clustering on such small scales, and additional clever observational ideas are undoubtedly  
waiting to be thought of. 
The SZ power spectrum is emerging as another promising cosmological observable,
sensitive the $k$-range near the cluster scale \cite{Seljak02}.

This avalanche of precision data offers an exciting challenge to theorists in the 
community: to raise the ambition level to making 
precision cosmology mean more than merely more apocryphal decimal  places, 
placing our understanding of the Universe on a solid foundation where the 
underlying physics has been tested rather than assumed.

The authors wish to thank Ang\'elica de Oliveira-Costa, 
Arthur Kosowsky, Uros Seljak and David Spergel for helpful comments.
This work was supported by 
NSF grants AST-0071213, AST-0134999, AST-0098606 and PHY-0116590,
NASA grants NAG5-9194 \& NAG5-11099,
and two Fellowships from the David and Lucile Packard Foundation.
MT is a Cottrell Scholar of Research Corporation.




\end{document}